\documentclass[%
reprint,
superscriptaddress,
amsmath,amssymb,longbibliography,
aps,
prl,
]{revtex4-1}
\pdfoutput=1
\usepackage{ulem}
\usepackage{hyperref}
\usepackage{graphicx}
\usepackage{amsmath}
\usepackage{color}
\usepackage{physics}
\usepackage{dsfont}
\usepackage{comment}
\usepackage{physics}
\usepackage{booktabs}
\usepackage{pifont}
\usepackage[T1]{fontenc}
\usepackage{makecell}
\usepackage{pdfpages} 
\usepackage{pgffor} 

\makeatletter
\AtBeginDocument{\let\LS@rot\@undefined}
\makeatother

\newif\ifarXiv
\arXivfalse

\definecolor{limegreen}{rgb}{0.2, 0.8, 0.2}
\definecolor{orange}{rgb}{1.0, 0.5, 0.0}

\definecolor{emerald}{rgb}{0.31, 0.78, 0.47}
\definecolor{blue(ncs)}{rgb}{0.0, 0.53, 0.74}

\def\red{\color{red}}

\hypersetup{
colorlinks=true,
linkcolor=magenta,
filecolor=blue,
citecolor=emerald,      
urlcolor =blue(ncs),
}
\def\k{{\mathbf{k}}}
\def\q{\mathbf{q}}

\begin{document}
\title{Quasiparticle Interference of Spin-Triplet Superconductors: Application to UTe$_2$}

\author{Hans Christiansen$^1$, Brian M. Andersen$^1$, P. J. Hirschfeld$^2$, and Andreas Kreisel}
\affiliation{Niels Bohr Institute, University of Copenhagen, DK-2100 Copenhagen, Denmark\\
$^2$Department of Physics, University of Florida, Gainesville, Florida 32611, USA}


\vskip 1cm

\begin{abstract}
Quasiparticle interference (QPI) obtained from scanning tunneling microscopy (STM) is a powerful method to help extract the pairing symmetry of unconventional superconductors. We examine the general properties of QPI on surfaces of spin-triplet          superconductors, where the properties of the $\vec d$-vector order parameter and topological surface bound states offer important differences from QPI on spin-singlet superconducting materials. We then apply the  theory to a model specific to UTe$_2$,   and compare the resulting QPI with recent STM measurements. We conclude that the two candidate Cooper pair instabilities $B_{2u}$ and $B_{3u}$ exhibit distinct features in the QPI intensity to discriminate these using the experimental data.
Characteristic features of the emergent topological surface states protected by chiral symmetry in general, and by mirror symmetries in the case of UTe$_2$, provide further unique signatures to help pinpointing the pairing symmetry channel in this material.
\end{abstract}
\maketitle

\textit{{\red Introduction.}} Spin-triplet superconductivity (SC) is a  fascinating state of matter with long sought-after properties for the development of future possible quantum technologies~\cite{Sarma2015,SatoFujimoto}. SC with spin-triplet Cooper pairs is richer than conventional $s$-wave SC in the sense that the condensed order allows for several different flavors, including for example chiral, helical, unitary, and non-unitary order parameters. These phases are symmetry-distinct and identified, for example, by the breaking of mirror or time-reversal symmetries (TRS). An important property of spin-triplet SC is the associated topological surface states (TSS) protected by bulk winding numbers or crystalline symmetries, depending on the flavor of the  triplet order, the Fermi surface topology, and the particular surface under consideration~\cite{Sato,Hsieh2012,Ishizuka,Henrik}. Indeed, such TSS and their potential robustness towards perturbations constitute the main desirable property of spin-triplet SC for applications. The existence of these surface states motivates the use of surface-sensitive experimental probes to access and manipulate the electronic surface properties. From a theoretical perspective, this requires the development of realistic surface theories able to capture the emergent TSS from the bulk Hamiltonian.

The heavy-fermion compound UTe$_2$ is currently under intense investigation due to intrinsic spin-triplet SC phase that may  be realized in its ground state~\cite{Ran,Aoki_2022,Lewin_2023}. This is deduced from critical magnetic fields larger than the Pauli-limiting field and the absence of a substantial Knight shift upon entering the SC phase~\cite{Ran,Aoki2019,Nakamine2019,Matsumura2023}. 
At present, the experimental status of the nature of SC in UTe$_2$ remains controversial, particularly with respect to the position of the point nodes and the possible realization of a chiral (non-unitary) condensate~\cite{Ishihara2023,Ajeesh,Suetsugu,Hayes_thermal,Theuss_2024,GuArXiv,Christiansen}. Thus, further experiments and theoretical studies are needed to determine the nature of  Cooper pairing in UTe$_2$.

Here, we focus on the technique of quasiparticle interference (QPI), an important probe for determining the gap structure of unconventional SC~\cite{QPI_Hofmann,QPI_Aynajian,QPI_Allan,QPI_sprau,QPI_Du,QPI_Kostin2,SeamusQPI}. We study the salient properties of QPI for spin-triplet SC in general, and UTe$_2$, in particular. Generically, QPI in triplet SC feature distinct properties compared to singlet SC, which needs to be taken into account for a proper interpretation of the QPI response. In the spin-singlet case, scattering takes place between states at momenta with the same or opposite signs of the order parameter, giving rise to interference effects that can be used to detect sign-changing singlet SC~\cite{QPI_Wang,Capriotti2003,Nunner2006,QPI_kreisel}. However, in a triplet SC, the $\vec d$-vector order parameter plays a crucial role for QPI and yields qualitatively different properties because of the vector nature of the order parameter. In particular, scattering can be suppressed or enhanced due to the direction of the $\vec d$-vector without direct connection to the magnitude of the $\vec d$-vector, which determines the energy of the Bogoliubov quasiparticles.  With various examples, we show how information on the structure of the ${\vec d}$-vector can be gleaned from the energy dependence of the scattering processes connecting van Hove points in the Bogoliubov spectrum on distinct contours of constant quasiparticle energy, provided the ${\vec d}$ vector is collinear.

Additionally, spin-triplet SC may be topological and host unique surface states important for the QPI. 
We demonstrate these properties of spin-triplet QPI by applying a realistic model of the SC state in UTe$_2$~\cite{Christiansen}. Specifically, we compute both bulk and surface Greens functions in the SC state and obtain the corresponding QPI from impurity scattering. For the particular case of UTe$_2$, it is important to address the experimentally relevant (0-11) cleave plane investigated by STM experiments. We find that the TSS offer unique signatures that can help pinpoint the pairing structure. From comparison to available experimental QPI data~\cite{SeamusQPI}, we conclude that the SC ground state of UTe$_2$ is likely to reside within the B$_{3u}$ symmetry channel.
\begin{figure}[t]
    \centering
    \includegraphics[width=\linewidth]{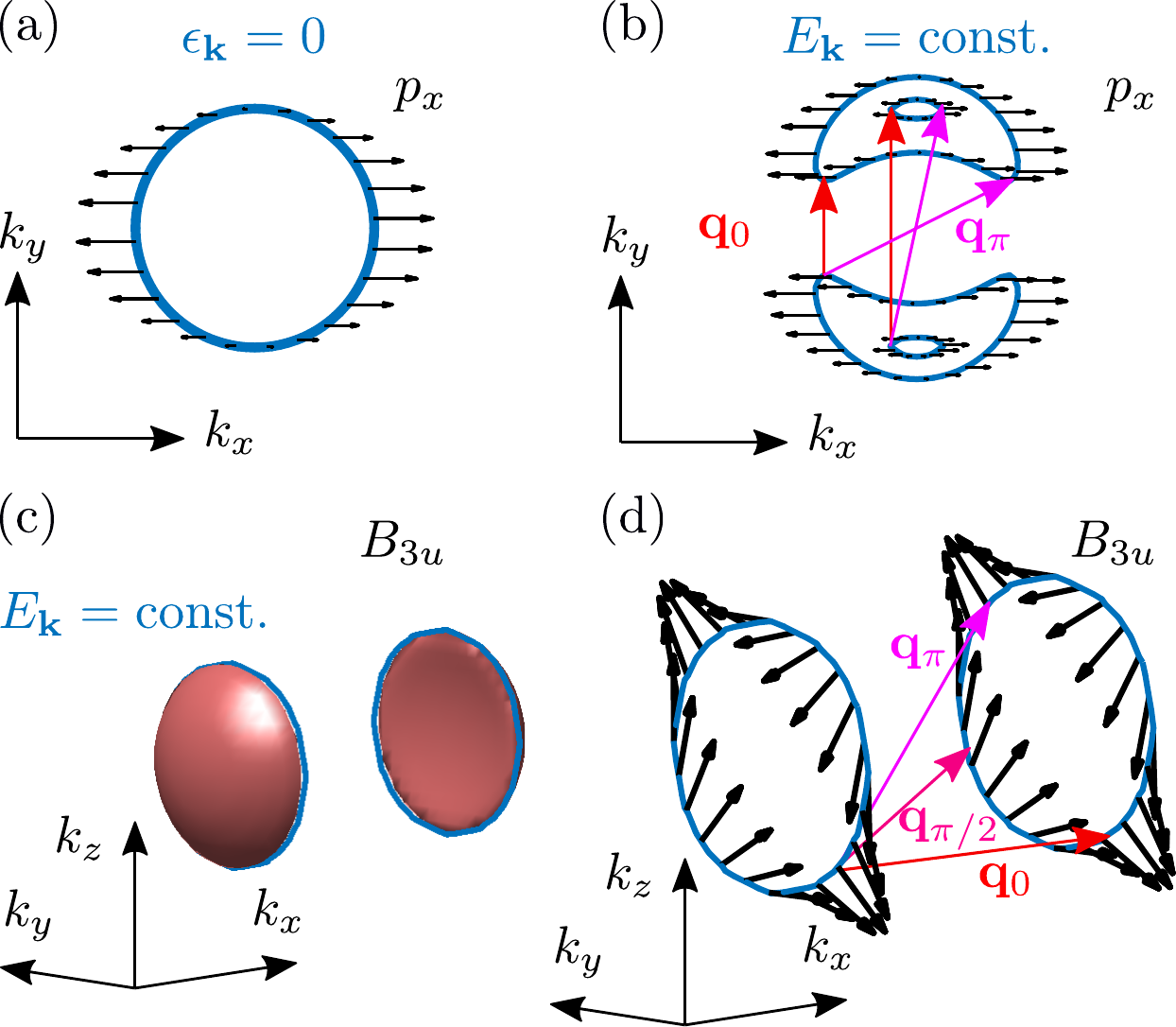}
    \caption{QPI in triplet superconductors (a) Fermi surface and $\vec d$ vector of a $p_x$ SC with $\vec d=p k_x\vec e_x$ in 2D. (b) Two contours of constant energy of the quasiparticle dispersion $E_{\mathbf k}
    $. For small energies, a ``banana'' close to the nodal point at $k_x=0$ occurs such that scattering processes with relative angle 0 between the $\vec d$-vectors (labeled by ${\bf q}_0$) and relative angle $\pi$ (labeled by ${\bf q}_\pi$) are difficult to resolve experimentally.
    At larger energies $\mathbf q_0$ and $\mathbf q_\pi$ can be more easily resolved and contain information about the relative direction of the $\vec d$-vector. (c) 3D analogon for a $B_{3u}$ state on a spherical Fermi surface, compare Fig. S3 in the SM. With the vector $\vec d_{B_{3u}}=(p_1k_xk_yk_z, p_2k_z, p_3k_y)$, there are point nodes at the $k_x$ axis such that at small energy the contours of constant energy form small ``lentils'' centered around the $k_x$ axis (red surface).
    Scattering at this energy is dominated by processes from the edges (blue lines) similar to the scattering processes of the tips of the ``bananas'' in two dimensions. (d) The $\vec d$-vector along these lines winds around such that there are scattering processes with all relative angles between the $\vec d$-vectors. Three example vectors with relative angle of the $\vec d$-vector of $0$, $\pi/2$ and $\pi$ are shown. At low energies, the ``lentils'' (and therefore the blue circles of large DOS) are small and $\mathbf q_0$ and $\mathbf q_\pi$ may not be resolvable experimentally, similar to the 2D case.}
    \label{fig:px_B3u}
\end{figure}

\textit{{\red QPI in spin-triplet superconductors.}} Before turning to a specific discussion of QPI in UTe$_2$, we provide a general study of the basic properties of QPI in triplet SC.
The starting point is a single-band model with the Bogoliubov-de Gennes (BdG) Hamiltonian
\begin{equation}\label{eq:bdgham}
   \check H({\mathbf k}) = \begin{pmatrix}
    H_N({\mathbf k})\sigma_0 && \Delta({\mathbf k}) \\
    \Delta^\dag({\mathbf k}) && -H_N^*(-{\mathbf k})\sigma_0
    \end{pmatrix} ,\
\end{equation}
in the Nambu basis  $\vec{c}_{\mathbf{k}} \equiv\left(c_{\mathbf{k} \uparrow}, c_{\mathbf{k} \downarrow}, c_{-\mathbf{k} \uparrow}^{\dagger}, c_{-\mathbf{k} \downarrow}^{\dagger}\right)$ with the normal state Hamiltonian $H_N({\mathbf k})=\xi_{\mathbf k}=\epsilon_{\mathbf k}-\mu$, where $\epsilon_{\mathbf k}$ is a dispersion and $\mu$ the chemical potential.
In terms of a $\vec d$-vector, the triplet order parameter is given by
\begin{equation}\label{eq:Deltatriplet}
 \Delta(\mathbf k)=(\vec d_{\mathbf k} \cdot \vec{\sigma})i\sigma_y\,,
\end{equation}
with quasiparticle energies $E_{\mathbf k}=\pm\sqrt {\xi_{\mathbf k}^2+|\vec d_{\mathbf k}|^2}$.
The momentum- and energy-resolved density modulations for scattering from $\mathbf{k}_F$ to $\mathbf{k}_F+\mathbf{q}_i$ from a non-magnetic impurity of strength $V_0$ are given by
\begin{equation}
    \delta \rho\left(\mathbf{k}_F, \mathbf{q}_i, \omega\right)= -2\frac{V_0}{\pi} \operatorname{Im}\left(\frac{(\omega+i \delta)^2-\vec {d}_{\mathbf{k}_{\mathrm{F}}} \cdot \vec {d}_{\mathbf{k}_{\mathrm{F}}+\q_i}^{*}}{\left[(\omega+i \delta)^2-|\vec{d}_{\mathbf{k}_{\mathrm{F}}}|^2\right]^2}\right),\label{eq_delta_rho}
\end{equation}
as derived in the Supplementary Material (SM).
Here, we have used the fact that dominant contributions occur from elastic scattering between two saddle points in the quasiparticle dispersion with $E_{\k}=E_{\k+\q_i}$. In addition, we have summed over spin, since we restrict our discussion to non-spin-polarized tunneling.
The contribution from a cross product term, $\vec d_{\k_F}\times\vec d_{\k_F+\q_i}^*$, which can be nonzero even in the unitary case, drops out, see SM.
The density modulations $\delta \rho$ in a singlet superconductor\cite{Hirschfeld2015} contain a term $\Delta_\k\Delta_{\k+\q}$, making the relative sign of the order parameter accessible in QPI. For the triplet superconductor, the corresponding expression,  Eq.~(\ref{eq_delta_rho}), exhibits the term $\vec {d}_{\mathbf{k}_{\mathrm{F}}} \cdot \vec {d}_{\mathbf{k}_{\mathrm{F}}+\q_i}^{*}$ which can take all values from $-|\vec {d}_{\mathbf{k}_{\mathrm{F}}}|^2$ to $+|\vec {d}_{\mathbf{k}_{\mathrm{F}}}|^2$.

For the simple $p_x$ superconductor, we have scattering processes labeled by $\mathbf q_\pi$ ($\mathbf q_0$) where this scalar product is negative (positive), see Fig. \ref{fig:px_B3u} (b). The antisymmetric density modulations $\rho^-(\omega)=\rho(\omega)-\rho(-\omega)$ as introduced in Ref.\cite{Hirschfeld2015}, exhibit no sign change (a sign change) from zero energy to the magnitude $|\vec {d}_{\mathbf{k}_{\mathrm{F}}}|$ for $\mathbf q_\pi$ ($\mathbf q_0$) similar to the singlet case with (without) sign change in the singlet order parameter  $\Delta_\k$.

This argument remains valid in three dimensions (3D) where for the $B_{3u}$ order parameter, the nodal points at zero energy evolve to contours of constant energy with $E_{\k}=\omega$ of ``lentil'' shape (Fig.~\ref{fig:px_B3u} (c)). Dominant scattering comes from the edges of the lentils (blue line) where the $\vec d$-vector winds as shown in Fig.~\ref{fig:px_B3u} (d).
Thus, we have three qualitative different behaviors of antisymmetric density modulations, $\rho^-(\omega)$: If the $\vec d$-vectors are perpendicular, as for example for $\mathbf q_{\pi/2}$, the QPI signal does not depend on the $\vec d$-vectors.
If the relative direction of $\vec {d}_{\mathbf{k}_{\mathrm{F}}}$ and $\vec {d}_{\mathbf{k}_{\mathrm{F}}+\q_i}^{*}$ is parallel (antiparallel), the corresponding antisymmetric density modulations, $\rho^-(\omega)$, exhibit a sign change (no sign change) from zero energy to the magnitude $|\vec {d}_{\mathbf{k}_{\mathrm{F}}}|$ similar to the singlet case~\cite{Hirschfeld2015}.
However, when considering scattering at low energies, i.e. for investigating the location of possible nodes, one integrates over momenta that connect $\bf k$-points with opposite directions of the $\vec d$-vector such that one averages over this quantity.
Consequently, it is not possible to deduce the relative direction of the $\vec d$-vector close to nodal points.
We note that the change of the direction of the $\vec d$-vector is imposed by symmetry, since the $\vec d$-vector changes sign across the nodal position, as shown in Fig.~\ref{fig:px_B3u}. In the SM, we discuss the $\vec d$-vector structure of other order parameters and scattering vectors $\mathbf q_0$ and $\mathbf q_\pi$.

\textit{{\red QPI in UTe$_2$.--}}
As a concrete timely example of QPI in a spin-triplet SC, we turn to the SC ground state of UTe$_2$. UTe$_2$ is a boby-centered orthorhombic material with D$_{2h}$ point group symmetry, allowing four symmetry-distinct odd-parity spin-triplet order parameters in the case of strong spin-orbit coupling (SOC): A$_u$, B$_{1u}$, B$_{2u}$, and B$_{3u}$. The BdG Hamiltonian and the order parameter are still given by Eq.~\eqref{eq:bdgham} and \eqref{eq:Deltatriplet}, but $H_N(\mathbf k)$ is now a $4\times 4$ matrix due to U or Te sublattices, similar to the formulation of the model found in Ref.~\cite{Christiansen}. The normal-state Hamiltonian $H_N(\mathbf k)$ is based on density functional theory (DFT) calculations and a four-band tight-binding fit matching recent quantum oscillation measurements~\cite{Theuss_2024,Eaton2024,Weinberger2024}. The four allowed SC odd-parity irreps restrict the possible pairing structures in sublattice and U/Te space and specify the final applied microscopic model for the SC phases of UTe$_2$. For all the details of model parameters and basis functions, we refer to Ref.~\cite{Christiansen} and the SM. STM on UTe$_2$ tunnels into the (0-11) cleave plane~\cite{GuArXiv,Jiao2020,SeamusQPI}. Thus, for comparison to experiments, theory necessarily needs to obtain the electronic states present at that particular surface. This is especially important for spin-triplet SC, where the odd-parity of the pair wavefunction $\Delta(-\mathbf k)=-\Delta(\mathbf k)$ tends to generate new low-energy Andreev states bound to the surfaces~\cite{buchholtz_1981,SatoFujimoto,Hsieh2012,Ishizuka,Geier_2020,Tei2023,Henrik,Christiansen}. The topological nature of these surface states depends on the topology of the Fermi surface. For the case of UTe$_2$ with open cylindrical Fermi sheets, they are weak and only protected by time-reversal and mirror symmetry along one direction of the (0-11) plane~\cite{Christiansen}. Nevertheless, {\it bona fide} massless Majorana surface modes emanating from time-reversal invariant momenta (TRIM) are generated in the present case, and hence relevant for the discussion of tunneling spectroscopy. Consequently, we apply an iterative scheme to obtain the bulk and surface Green's functions, $G_{s}(\mathbf k^\parallel,\omega)$ and $G_{b}(\mathbf k^\parallel,\omega)$, associated with the (0-11) surface. Here, $\mathbf{k}^\parallel$ refers to the momentum parallel to the surface i.e. $\mathbf{k}^\parallel=k_x\vec m_x + k_{c^*}\vec m_{c^*}$, with $\vec m_x = (1/a,0,0)$ and $\vec m_{c^*}=(0,1/b,1/c)$. For a detailed discussion of the Green's functions and the topological aspects of the surface states, we refer to the SM and Refs.~\cite{sancho_highly_1985,Tei2023,Henrik,Christiansen}.

Having obtained the homogeneous Green's functions, we turn to QPI signals induced by impurities on the \mbox{(0-11)} surface. As shown in the SM, the generalization of the density modulation from an impurity with strength $V_0$ in Eq.~\eqref{eq_delta_rho} for the multi-band case is
\begin{equation}\label{eq:deltarhoalphabeta}
    \delta \rho_{\beta}(\mathbf{q}^{\parallel},\omega) = -\frac{V_0}{2\pi i}( f_{\beta}(\mathbf q^{\parallel},\omega) - f^*_{\beta}(-\mathbf q^{\parallel},\omega))~,
\end{equation}
with
\begin{equation}
f_{\beta}(\mathbf q^{\parallel},\omega) = \sum_{\mathbf k^{\parallel}} \mathrm{Tr}\qty(\tau_e G(\mathbf k^{\parallel}+\mathbf q^{\parallel},\omega)\tau_3P_\beta  G(\mathbf k^{\parallel},\omega)),\label{eq:fQPI}
\end{equation}
where $\tau_3$ is the third Pauli matrix in Nambu space, $\tau_e = (\tau_0 + \tau_3)/2$ projects onto the electronic sector, and $P_{\beta}$ is a projector onto U/Te and one of the sublattice degrees of freedom. In Eq.~(\ref{eq:fQPI}), $G(\mathbf k^{\parallel},\omega)$ refers to either the bulk or surface Green's function, both being relevant depending on which states are probed by the tunneling processes. The total density modulation is computed as $\delta\rho(\mathbf{q}^{\parallel},\omega) = \sum_{\beta}\abs{\rho_\beta(\mathbf q^{\parallel},\omega)}$ where the absolute value arises from the assumption of homogeneously distributed impurities. This means that we sum over contributions from impurities on both Te and U sites with equal impurity potential. See SM for the QPI signal arising specifically from U or Te disorder. Finally we note that as discussed above, scattering from node to node does not generate signatures in $\rho^-(\omega)$ that determine the direction of the $\vec d$-vector. The presence of low-energy scattering in $\delta \rho_\beta(\mathbf{q}^\parallel,\omega)$, however, can reveal the relative positions of the nodes, and constrain the SC order parameters from available experimental data, which we pursue below. 
\begin{figure}[t]
    \centering
    \includegraphics[width=\linewidth]{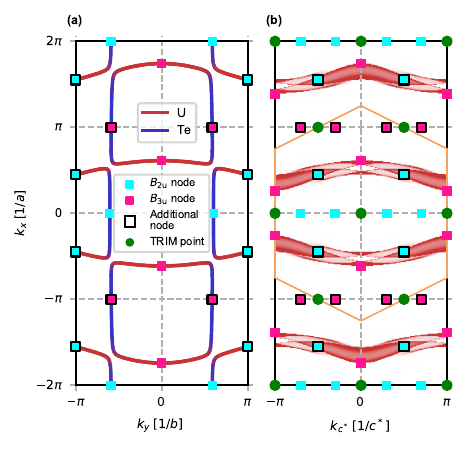}
    \caption{(a) Fermi surface at $k_z=0$ with red (blue) bulk bands dominated by U (Te) orbital content. (b) (0-11) spectral function in the normal state, arising mainly from the U bands. The locations of symmetry-imposed and additional nodes for both B$_{2u}$ and B$_{3u}$ have been indicated in both panels, along with the TRIM points relevant for the TSS. The orange lines in (b) indicate the surface Brillouin zone.}
    \label{fig:setstage}
\end{figure}

\begin{figure*}[t]
    \centering
    \includegraphics[width=\linewidth]{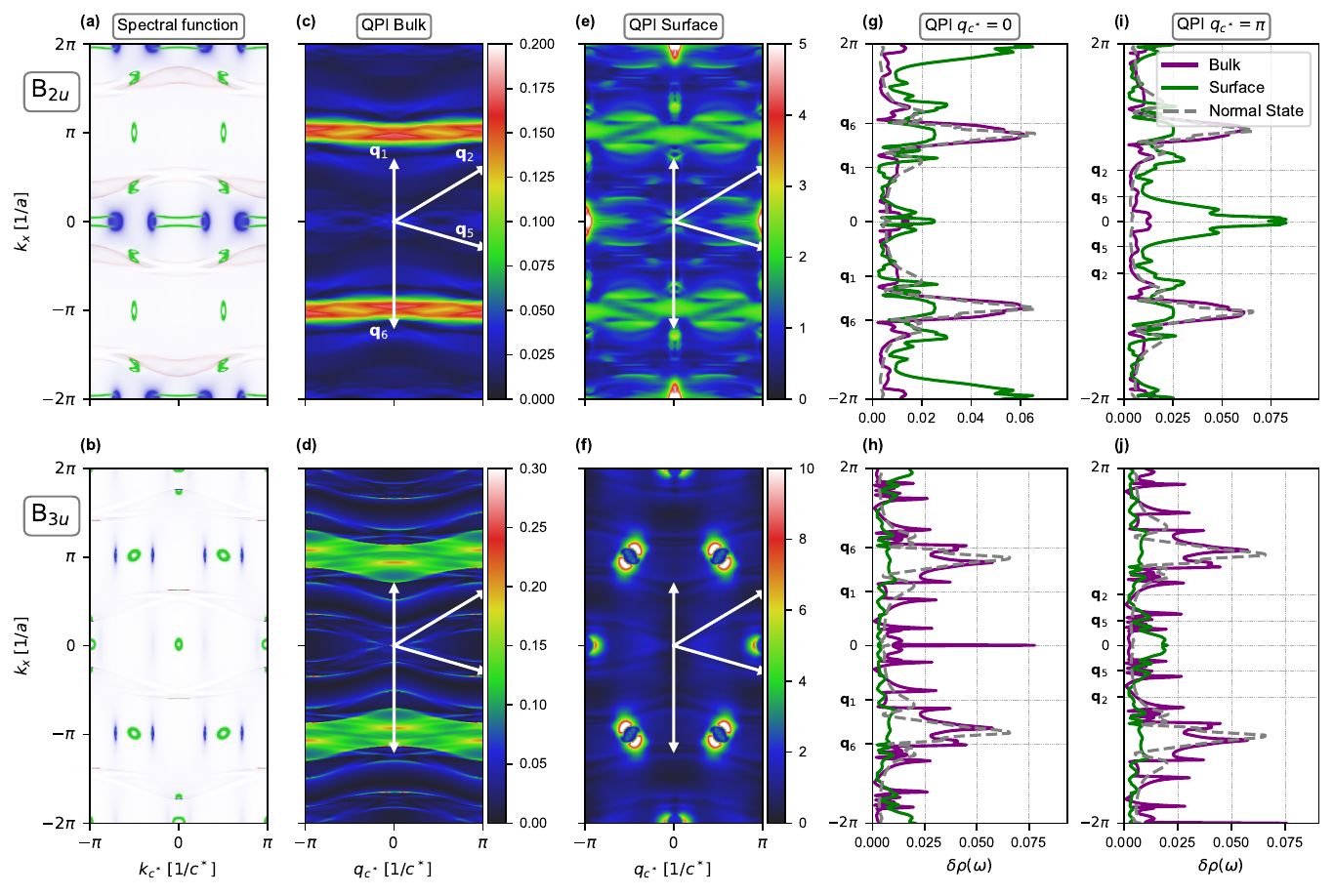}
    \caption{QPI signal $\delta\rho(\mathbf{q}^{\parallel},\omega)$ and spectral functions $A_{s/b}(\mathbf{q}^\parallel,\omega)=-\frac{1}{\pi}\mathrm{Im}G_{s/b}(\mathbf{q}^{\parallel},\omega)$ on the (0-11) surface of UTe$_2$ in the B$_{2u}$ and B$_{3u}$ phases. (a-b) The bulk U (red) and Te (blue) spectral functions along with the surface states (green). (c,d) [e,f] QPI versus surface momentum at $\omega=0.05\Delta_0$ including the surface-projected bulk [surface] states. Characteristic scattering vectors $\mathbf q_i$ as seen by the STM experiment in Ref.~\cite{SeamusQPI} are indicated by the white arrows. (g,h) [(i,j)] Momentum cuts of the respective panels (c,d) [e,f] at $q_{c^*}=0, \pi$. In panels (g-j) we have also indicated the locations of $\mathbf q_i$. As seen, only the B$_{3u}$ pairing symmetry features significant enhancements at $\mathbf q_1$ and $\mathbf q_5$ compared to the normal state QPI response. In panels (c-f) the  plots are in units of $V_0/{\mathrm{(eV)}^2}$ while in panels (g-j) the density modulations have been normalized such that $\sum_{\mathbf{q}^\parallel}\delta\rho(\mathbf{q^\parallel},\omega)=1$.}
    \label{fig:mainQPI}
\end{figure*}

To set the stage for the following discussion on QPI in UTe$_2$, we show in Fig.~\ref{fig:setstage}(a) the bulk bands at $k_z=0$ with nodes highlighted for B$_{2u}$ (B$_{3u}$). The symmetry-imposed nodes are located along the $k_y$ ($k_x$) axis, whereas additional nodes are present at other locations of the Fermi surface, as shown in Fig.~\ref{fig:setstage} and elaborated in the SM~\cite{Christiansen}. At low energies the quasiparticle scattering is restricted to connect nodal regions, giving rise to characteristic scattering vectors in the QPI signal.
Figure~\ref{fig:setstage}(b) displays the bands projected to the experimentally relevant (0-11) cleave plane, and the positions of the nodes and the TRIM points on that surface. As seen, only the U-dominated states lead to distinct bands on the surface.
By contrast, the Te bands produce an almost constant background weight due to the angle of the cleave plane and the dispersion with large Fermi velocity of those bands, see SM for further details of the (0-11) coordinates and surface Brillouin zone.
In the presence of impurities, a characteristic normal state QPI pattern  emerges from scattering between the bands shown in Fig.~\ref{fig:setstage}(b). In the SC phase, low-energy QPI is strongly restricted due to the gap, but additional complexity arises for triplet SC due to the structure of the $\vec d$-vector (see above) and the emergence of TSS.

Figure~\ref{fig:mainQPI}(a,b) display the spectral functions of both the bulk and surface for the case of B$_{2u}$ and B$_{3u}$ SC, respectively.
As seen from comparison to Fig.~\ref{fig:setstage}(b), spectral intensity exists both near the surface-projected nodes from the bulk bands and at the TRIM points due to the emergent TSS. Disorder generates scattering between these quasiparticle states, producing a spectacle of interference patterns shown in Fig.~\ref{fig:mainQPI}(c-f). The bulk signal in Fig.~\ref{fig:mainQPI}(c,d) arises mainly from the dispersion of the bulk U bands, while the surface signal in Fig.~\ref{fig:mainQPI}(e,f) originates from the TSS which have predominant weight on the Te sites. In the Born limit, the QPI intensity is directly proportional to the impurity potential, implying that from tuning the disorder, one might be able to interpolate between the bulk and surface dominated QPI signal, as discussed in the SM. 
QPI linecuts at $k_{c^*}=0,\pi$ are shown for both the normal state and the B$_{2u}$ and B$_{3u}$ SC cases in Figs.~\ref{fig:mainQPI}(g-j). From the QPI data shown in Fig.~\ref{fig:mainQPI}, we draw two main conclusions: 1) B$_{3u}$ is the most likely pairing channel realized in UTe$_2$, and 2) distinctive QPI patterns from TSS offer additional means to distinguish between B$_{2u}$ and B$_{3u}$ SC. 

Conclusion 1) results from the fact that the enhanced scattering channels dubbed $\mathbf q_1$ and $\mathbf q_5$ in Ref.~\cite{SeamusQPI}, as shown explicitly in Fig.~\ref{fig:mainQPI}, are generically only significantly enhanced compared to the normal state in the B$_{3u}$ phase, see also Ref.~\cite{Crepieux2025}. This is because only for B$_{3u}$, these scattering vectors arise from Bogoliubov quasiparticle scattering between symmetry-imposed nodes, see Fig.~\ref{fig:setstage}(b). In the SM, we discuss a special (fine-tuned) case where B$_{2u}$ exhibits scattering at $\mathbf q_1$ and $\mathbf q_5$ from the additional nodes. Even in that case, however, the conclusion remains that only B$_{3u}$ features enhanced QPI response at $\mathbf q_1$ and $\mathbf q_5$ compared to the normal state. Conclusion 2) arises from the fact that only the B$_{2u}$ state features Majorana flatband TSS whereas B$_{3u}$ supports Majorana-Dirac TSS dispersions, see Fig.~\ref{fig:mainQPI}(e,f) and Fig.~\ref{fig:mainQPI}(g-j). The origin of this qualitative difference can be traced to the different properties of B$_{2u}$ and B$_{3u}$ under mirror along $k_x$~\cite{Christiansen}. For the resulting QPI patterns, this leads to distinct flat bands (cones) of scattering intensity for B$_{2u}$ and B$_{3u}$, respectively, as seen from Fig.~\ref{fig:mainQPI}(e,f). The low-energy quasiparticles close to the TRIM point, $\mathbf{k}_n^\parallel$, can be described by an effective Hamiltonian $H_{n,\mathrm{eff}}(\delta \mathbf{k}^\parallel)= \vec d_n(\delta \mathbf{k}^\parallel)\cdot \vec \Gamma$ using the two zero-energy states at the TRIM point as basis states and the vector of Pauli matrices $\vec \Gamma$.
In the appropriate basis, the spin polarization of the TSS determines their effective ${\vec d_n}$-vector which, in turn, strongly affects the surface QPI. The difference to the simple ${\vec d}$-vectors discussed in Fig.~\ref{fig:px_B3u} lies in the structure of the surface state.
First, there is no scattering allowed within the same Dirac cone due to the Kramers degeneracy and the chiral anti-symmetry at the TRIM points.
Second, the winding of the ${\vec d}$-vector near the TRIM points determines the QPI intensity in Fig.~\ref{fig:mainQPI}(e,f) such that nodes in the scattering amplitude are expected if the states at the two relevant Dirac cones wind in the same direction.
We explain in detail in the SM how the matrix elements of the impurity potential between the basis states at the TRIM point and the winding of the ${\vec d}$-vector enters the calculation of the scattering intensity.

In comparison with the experimental QPI data available on UTe$_2$~\cite{SeamusQPI}, no distinctive TSS features stand out near the TRIM points. This can be explained if the QPI signal is dominated by scattering from U impurities, since visibility of the TSS would need significant scattering from Te sites. Alternatively, it may be due to the fact that the crystalline TSS in the present case are only protected by TRS and one mirror, making their distinctive features fragile to disorder on the surface. A large resulting featureless spectral weight from these surface states may be related to the large background conductance seen experimentally~\cite{GuArXiv,SeamusQPI,Yangvortex2025,Yinvortex2025,Sharmavortex2025}. Additionally, UTe$_2$ is known to host a weak surface  charge density wave (CDW) which is manifested by the nondispersive QPI peaks $\mathbf q_2$, $\mathbf q_6$, and possibly $\mathbf q_1$~\cite{Aishwarya2023,Gu2023,Kengle2024,Suderow2025}. Since the model applied in this work already contains enhanced scattering at those wave vectors in the normal state, the surface CDW may arise from favorable nesting properties at those momenta transfer. It constitutes an interesting future study to include the weak surface CDW explicitly in the calculations of the QPI. We do not, however, anticipate this to qualitatively alter our conclusions of the QPI response in the superconducting state.

\textit{{\red Conclusions.--}} We have provided a general study of QPI in spin-triplet superconductors and discussed the differences from the spin-singlet case arising from the vectorial nature of the triplet order parameter. We focused on the  timely example of UTe$_2$, and computed the QPI response of this material on the experimentally relevant (0-11) cleave plane. From comparison to currently available experimental SC QPI data~\cite{SeamusQPI}, we conclude that B$_{3u}$ pairing appears most likely as the ground state symmetry of SC in UTe$_2$. Distinct topological surface states in the cases of B$_{2u}$ and B$_{3u}$ pairing offer an additional possibility for distinguishing these phases in future STM experiments on pristine surfaces.  
\begin{acknowledgments}
\textit{{\red Acknowledgements.--}} 
We acknowledge discussions with J.~C.~Séamus Davis and Max Geier. H.~C. acknow\-ledges support from the Novo Nordisk Foundation grant NNF20OC0060019. A.~K. acknowledges support by the Danish National Committee for Research Infrastructure (NUFI) through the ESS-Lighthouse Q-MAT. 
\end{acknowledgments}

\bibliography{Refs}

\end{document}


\title{Supplemental Information: Quasiparticle Interference of Spin-Triplet Superconductors: Application to UTe$_2$}
\author{Hans Christiansen$^1$, Brian M. Andersen$^1$, P. J. Hirschfeld$^2$, and Andreas Kreisel$^1$}
\affiliation{$^1$Niels Bohr Institute, University of Copenhagen, DK-2100 Copenhagen, Denmark\\
$^2$Department of Physics, University of Florida, Gainesville, Florida 32611, USA}
\date{\today}
\begin{abstract}
This supplemental information presents details of the theoretical methods used and provides the parameters employed in the calculations.
\end{abstract}
\maketitle

\section{Spin-triplet superconductor: Quasiparticle interference in the Born approximation}

We start from the BdG Hamiltonian
\begin{equation}\label{eq:bdgham}
   \check H({\mathbf k}) = \begin{pmatrix}
    H_N({\mathbf k})\sigma_0 && \Delta({\mathbf k}) \\
    \Delta^\dag({\mathbf k}) && -H_N^*(-{\mathbf k})\sigma_0
    \end{pmatrix} ,\
\end{equation}
in the Nambu basis  $\vec{c}_{\mathbf{k}} \equiv\left(c_{\mathbf{k} \uparrow}, c_{\mathbf{k} \downarrow}, c_{-\mathbf{k} \uparrow}^{\dagger}, c_{-\mathbf{k} \downarrow}^{\dagger}\right)$, with $H_N({\mathbf k})=\xi_{\mathbf k}=\epsilon_{\mathbf k}-\mu$ where $\epsilon_{\mathbf k}$ is a dispersion and $\mu$ the chemical potential. We construct the
bare Green's function ${\check{G}}_0(\k,\tau)$$\equiv \langle {{\cal T}}_\tau \vec{c}_{\mathbf{k}}(\tau) \vec{c}_{\mathbf{k}}^\dagger\rangle$ and calculate it in frequency space as
\begin{equation}
\check{G}_0(\k,\omega)=(\check{1} \omega-\check{\mathbf{H}})^{-1}=\left(\begin{array}{ll}
\hat{G}_{11} & \hat{G}_{12} \\
\hat{G}_{21} & \hat{G}_{22}
\end{array}\right),
\end{equation}
where $\omega$ is the quasiparticle energy. The Matsubara Green's function can be obtained by $\omega \rightarrow i \omega_n$.  The $2 \times 2$ matrices in  spin space can be explicitly calculated as
\begin{align}
\hat{G}_{11}&=\frac{(\omega+\xi)}{D}\left[\left(\omega^2-\xi^2-|\vec {d}|^2\right) \sigma_0+ \vec{q} \cdot \vec{\sigma}\right], \\
\hat{G}_{12}&=\left[\left(\omega^2-\xi^2-|\vec {d}|^2\right) \sigma_0+ \vec{q} \cdot \vec{\sigma}\right] \frac{i (\vec{d} \cdot \vec {\sigma}) \sigma_y}{D},
\\
\hat{G}_{21}&=-\left[\left(\omega^2-\xi^2-\Delta_0^2|\vec{d}|^2\right) \sigma_0+ \vec{q} \cdot \vec{\sigma}^T\right] \frac{ i \sigma_y\left(\vec{d}^{*} \cdot \vec{\sigma}\right)}{D}, \\
\hat{G}_{22}&=\frac{(\omega-\xi)}{D}\left[\left(\omega^2-\xi^2-|\vec{d}|^2\right) \sigma_0+ \vec{q} \cdot \vec{\sigma}^T\right],
\end{align}
where we introduced the notation $\vec{q}=i\left(\vec {d} \times \vec {d}^*\right)$ and
wrote triplet order parameter in terms of a $\vec d$-vector as
\begin{equation}
 \Delta(\mathbf k)=(\vec d_{\mathbf k} \cdot \vec{\sigma})i\sigma_y\;.\label{eq_delta}
\end{equation}
We dropped the index for the momentum dependence and introduced the denominator from the matrix inversion given by
\begin{align}
D
=\left(\xi^2-\omega^2+\Delta_{+}^2\right)\left(\xi^2-\omega^2+\Delta_{-}^2\right).
\end{align}
The quantity $\Delta_{ \pm}^2=\left(| \vec{d} |^2 \pm | \vec{q} |\right)$ appears in the quasiparticle energies; for the unitary case, $\vec{q}=0$; there is only a single energy scale.
For TRSB non-unitary states, $\vec{q} \neq 0$, that can occur as a mixture of multiple irreducible representations, the excitation energies become non-degenerate.
$\vec q$ can be interpreted as the spin moment of the Cooper pairs and the average of $\vec{q}$ over the Fermi surface vanishes in a antiferromagnetic state and does not vanish for a ferromagnetic state.

Restricting to the unitary case, the above expressions simplify to the components of the Green functions as
\begin{eqnarray}
\hat{G}_{11}=\frac{(\omega+\xi)}{\tilde D} \sigma_0, \\
\hat{G}_{12}=\frac{i (\vec {d} \cdot \vec {\sigma}) \sigma_y}{\tilde D} \\
\hat{G}_{21}=- \frac{ i \sigma_y\left(\vec {d}^{*} \cdot \vec {\sigma}\right)}{\tilde D} \\
\hat{G}_{22}=\frac{(\omega-\xi)}{\tilde D} \sigma_0
\end{eqnarray}
where the denominator is now
\begin{equation}\tilde D=\xi^2-\omega^2+|\vec d|^2 .
\end{equation}

\section{Comparison between singlet and triplet QPI: HAEM}

A possible phase-sensitive probe for the superconducting order parameter has been introduced in Ref. \cite{Hirschfeld2015} (HAEM) where it was demonstrated that the antisymmetric density of states $\delta\rho^{-}(\omega)\equiv\delta\rho^{\mathrm{inter}}(\omega)-\delta\rho^{\mathrm{inter}}(-\omega)$ has signatures of the relative sign of the (singlet) order parameter $\Delta_{\mathbf k}$ between different bands.
Here, we repeat these arguments and compare to the case of a triplet superconductor.
Focusing on the leading order in the impurity potential $\hat V=V_{0}\tau_3\sigma_0$, i.e. Born approximation, we can calculate the modulations of the density of states as
\begin{align}
\delta \rho_s
(\mathbf{q}, \omega)=&-\frac{1}{2i\pi} \sum_{\mathbf{k}}
(\operatorname{Tr}\{ P_{e,s}\hat{G}_{\mathbf{k}}(\omega) \hat{V} \hat{G}_{\mathbf{k}-\mathbf{q}}(\omega)\}\notag\\
& -\operatorname{Tr}\{ P_{e,s}\hat{G}_{\mathbf{k}}(\omega) \hat{V} \hat{G}_{\mathbf{k}+\mathbf{q}}(\omega)\}^*)\,,
\label{eq_density_mod}
\end{align}
where $P_{e,s}$ projects to the particle part of spin $s$.
Note that the experimentally accessible QPI is obtained from a Fourier transform of the real-space conductance maps which can be calculated by $\delta \rho_s
(\mathbf{r}, \omega)=-\frac {1}{\pi} \mathrm{Im} \operatorname{Tr}\{ P_{e,s} \hat{G}_{\mathbf{r}}(\omega) \hat{V} \hat{G}_{-\mathbf{r}}(\omega)\}$, i.e. the imaginary part has to be taken before the Fourier transform, leading to the two terms in Eq. (\ref{eq_density_mod}).
The original idea of HAEM for singlet superconductors was based on calculating the density modulations  $\delta\rho^{\mathrm{inter}}(\omega)\equiv \sum_{\q\sim \q_0} \delta\rho(\q,\omega)$ at momentum transfer $\q_0$ for an electronic structure that consists of two pockets separated by his momentum and then examining  $\delta\rho^{-}(\omega)\equiv\delta\rho^{\mathrm{inter}}(\omega)-\delta\rho^{\mathrm{inter}}(-\omega)$, which had ``even'' or ``odd'' properties as a function of $\omega$ depending on whether $\q_0$ connected gaps of different or opposite sign, respectively.
This idea was expanded by the work of Böker et al. \cite{Boeker2020} by integrating around $\q_i$ to get a momentum resolved HAEM with the result
\begin{equation}
    \delta \rho\left(\mathbf{k}_F, \mathbf{q}_i, \omega\right)=-\frac{V_0}{\pi} \operatorname{Im}\left(\frac{(\omega+i \delta)^2-\Delta_{\mathbf{k}_F} \Delta_{\mathbf{k}_F+\mathbf{q}_i}^*}{\left[(\omega+i \delta)^2-\left|\Delta_{\mathbf{k}_F}\right|^2\right]^2}\right).\label{eq:dens_mod}
\end{equation}
It was tested on a d-wave superconductor where the expected dominating scattering vectors come from saddle points in the quasiparticle dispersion which appear on 8 momenta in the Brillouin zone. In the so-called octet model~\cite{QPI_Wang}, the HAEM technique can be used to determine which momentum transfer connecting the 8 points corresponds to sign-changing scattering processes by integrating over a small range of $\mathbf{q}$ around each octet spot individually.
In particular, one can analytically calculate the antisymmetric density modulation $\delta \rho^{-}(\mathbf{k}, \mathbf{q}, \omega)$ ``pointwise'' (i.e., at specific values of $\mathbf{q}$ ) and evaluate it at the saddle point momenta (banana tips) where the first momentum is at the Fermi surface, $\mathbf{k}=\mathbf{k}_F$ and the momentum transfer at one of the octet vectors, $\mathbf{q}=\mathbf{q}_i$. Exactly at these points, the normal state dispersion vanishes, $\epsilon_{\mathbf{k}_F}=\epsilon_{\mathbf{k}_F+\mathbf{q}_i}=0$, and the order parameter magnitude is identical, $\Delta_{\mathbf{k}_F+\mathbf{q}_i}= \pm \Delta_{\mathbf{k}_F}$.~\cite{Hirschfeld2015,Boeker2020,Martiny,Nunner2006,Shinibali},

In a triplet superconductor, we still have saddle points of the same type at the Fermi momenta, but the $\vec d$-vector is not necessarily collinear any more, i.e.  $\vec d_{\mathbf{k}_F+\mathbf{q}_i}= \pm \vec d_{\mathbf{k}_F}$ does not hold which leads to additional complications.
Continuing to calculate the density modulations given by Eq. (\ref{eq_density_mod})
we find
\begin{widetext}

\begin{align}
\delta \rho_{\uparrow,\downarrow}\left(\mathbf{k}_F, \mathbf{q}_i, \omega\right)
=
-\frac{V_0}{2i\pi} \left(\frac{(\omega+i \delta)^2-(\vec d_{\k_F}\cdot{\vec d}_{\k_F-\q_i}^*)\pm i(\vec d_{\k_F}\times\vec d_{\k_F-\q_i}^*)_z}{\left[(\omega+i \delta)^2-\left|\vec d_{\mathbf{k}_F}\right|^2\right]^2 }
-\frac{(\omega-i \delta)^2-(\vec d_{\k_F}^*\cdot{\vec d}_{\k_F+\q_i})\mp i(\vec d_{\k_F}^*\times\vec d_{\k_F+\q_i})_z}{\left[(\omega-i \delta)^2-\left|\vec d_{\mathbf{k}_F}\right|^2\right]^2}
\right)\label{eq_delta_rho1}
\end{align}
\end{widetext}
using
$(\vec{a} \cdot \vec{\sigma})(\vec{b} \cdot \vec{\sigma})=(\vec{a} \cdot \vec{b})\sigma_0+i(\vec{a} \times \vec{b}) \cdot \vec{\sigma}$.
Note that the cross product term does not vanish even for unitary case where  ${\vec d_{\k}}\times{\vec d^*_{\k}}=0$ vanishes only for the same momenta. We use that the $\vec d$-vector is odd parity $\vec d_{\mathbf k}=-\vec d_{-\mathbf k}$ to simplify Eq. (\ref{eq_delta_rho1}).
Focusing on the spin-summed result (that is mostly available in non-spin polarized STM experiments), $\delta\rho =\delta\rho_\uparrow+\delta\rho_\downarrow$, we obtain by using that the magnitude of the $\vec d$-vector is identical, $|\vec d_{\k_F}|= |\vec d_{\k_F+\q_i}|$, for elastic scattering with $E_{\k}=E_{\k+\q_i}$ from and two saddle points,
\begin{equation}
    \delta \rho\left(\mathbf{k}_F, \mathbf{q}_i, \omega\right)= -\frac{2V_0}{\pi} \operatorname{Im}\left(\frac{(\omega+i \delta)^2-\bigl(\vec{d}_{\mathbf{k}_{\mathbf{F}}} \cdot \vec{d}_{\mathbf{k}_{\mathbf{F}}+\q_i}^{*}\bigr)}{\bigl[(\omega+i \delta)^2-\bigl|\vec {d}_{\mathbf{k}_{\mathbf{F}}}\bigr|^2\bigr]^2}\right).\label{eq_delta_rho}
\end{equation}

The relative orientation of $\vec d_{\k_F}$ and $\vec d_{\k_F+\q_i}$ (when allowing $\q_i$ to vary slightly) yields the fingerprint in the antisymmetric $\delta \rho^-(\omega)$ where one has to distinguish three cases in general. If the momentum transfer relates opposite-parity momenta (up to a reciprocal vector $\mathbf G$),  $\k+\q_i= -\k +{\bf G}$, the $\vec d$-vectors are anti-parallel due to the odd parity of the order parameter, $\vec d_{\k_F+\q_i}=-\vec d_{\k_F}$.
Then $\delta \rho^-(\omega)$ will exhibit a symmetric form as worked out from a similar expression for the density modulations (Eq.(\ref{eq_delta_rho})) in the singlet case. Points at high symmetry where $\vec d_{\k_F+\q_i}= \vec d_{\k_F}$ yield an antisymmetric form of $\delta \rho^-(\omega)$ as in the singlet case as well. Should the $\vec d$-vectors not have a fixed relative direction when varying $\q_i$ slightly, there is not a clear signature of sign-changing or non-sign changing in the frequency dependence of $\delta \rho^-(\omega)$. This can happen in two ways: First, the $\vec d$-vector with multiple components might rotate around the extremal edge of the surface of constant quasiparticle energy or second, in the regime where $|\vec d_{\k_F}|$ is small near nodal points,  the direction of the $\vec d$-vector has to flip direction if one of its components flips sign  when one of the components of $\k$ flips.  Very close to the node, any experiment will average over both  directions, and the $\rho^-$ signal will cancel.  At this point, we note that the identification of the relative direction of the $\vec d$-vector might still work for systems with distinct Fermi surface pockets that do not hit the nodes of the odd parity order parameter as for example discussed for WTe$_2$\cite{Jahin2023}.

\subsection{Examples for QPI in triplet superconductors}

\subsubsection{2D $p_x$ superconductor}

\begin{figure}[tb]
    \centering
 \includegraphics[width=\linewidth]{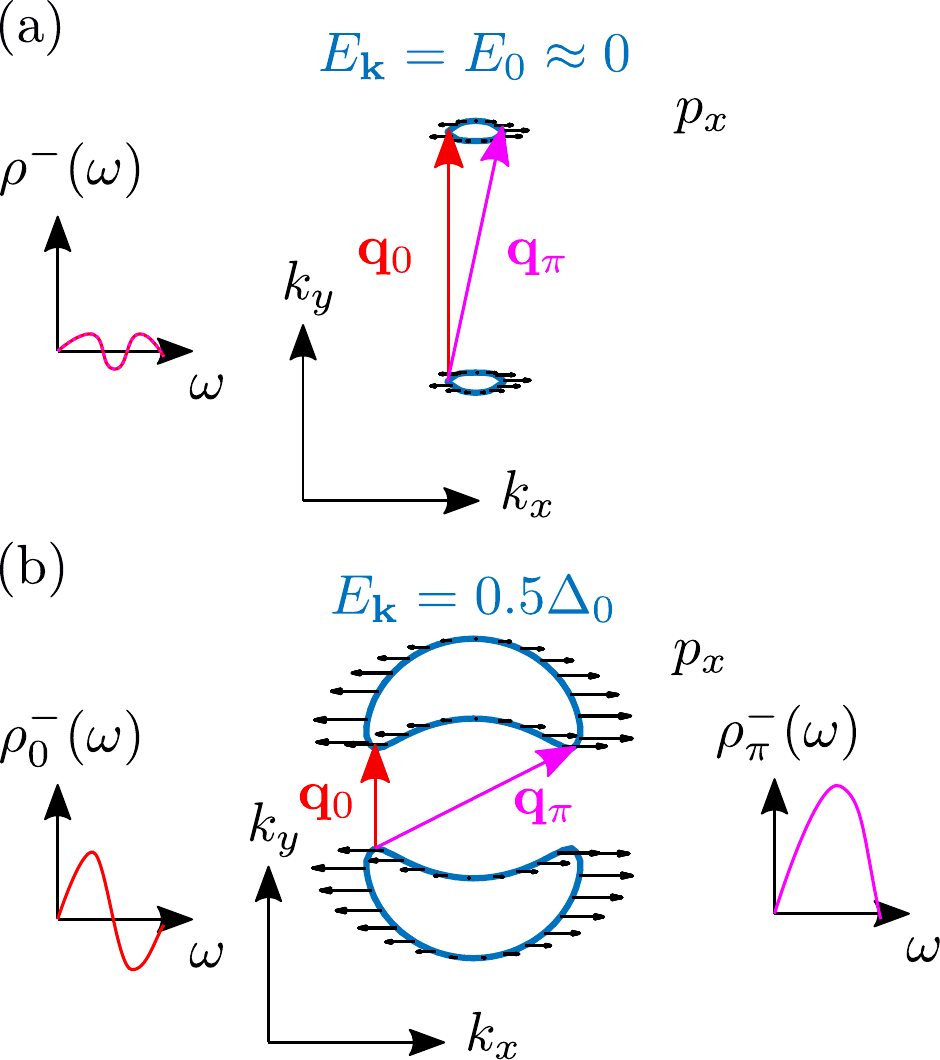}
    \caption{Contours of constant quasiparticle energy in a $p_x$-wave superconductor and QPI: (a) At small energies, the two types of scattering vectors for the same direction of the $d$-vector $\q_0(\omega)$ and opposite direction $\q_{\pi}(\omega)$ are almost indistinguishable, so no strong signature in $\rho^-(\omega)$ is expected due to averaging. (b) for larger energy, the Fermi momenta with $\k_F$ and $-\k_F$ have opposite directions of the $d$-vector by odd parity of the order parameter, so $\rho^-(\omega,\q_0(\omega))$ should  display an $\omega$-dependence resembling the schematic in the left panel, while $\rho^-(\omega,\q_\pi(\omega)$  exhibits no sign change (right).
    }
    \label{fig:px}
\end{figure}
Focusing on the low-energy quasiparticle interference where from Fermi's golden rule, only quasiparticles close to the nodes can scatter as also described by the octet-type models at low energies. For a simple $p_x$ superconductor in two dimensions with small Fermi surfaces, there are two types of scattering vectors around a node. In the case of small $\omega$ where the bananas are also small, the two vectors $\q_0$ and $\q_\pi$ are hard to distinguish and $\delta \rho^-(\omega)$ at these two vectors have opposite behavior with respect to the energy dependence, i.e. the behavior cannot be used to detect the relative direction of the $\vec d$-vectors. For larger energies, the $\q$ vectors are more separated and the antisymmetric density of states behaves analog to the singlet case if followed in momentum space according to $\q(\omega)$, see Fig. \ref{fig:px}(a,b) such that in this example of a two dimensional superconductor with collinear $\vec d$-vector, the method can be used to find the relative directions of the triplet order parameter.

\subsubsection{ABM state}
For illustration purposes of a three dimensional system, we start with a well known example of a triplet superconductor as realized in the A-phase of $^3$He, the Anderson-Brinkmann-Morel (ABM) state which has fixed $\vec d_{\k}$ direction but not magnitude,
\begin{equation}
\vec d_{\k}=\Delta_0 (\hat k_x+i\hat k_y){\vec e_z}.
\end{equation}
The equation $\omega=E_{\k}=\sqrt{{\xi_{\k}}^2+|\vec d_{\k}|^2}$ turns out have contours of constant quasiparticle energy (CCE) that look like meniscus lenses (not unlike taking a 2D octet model banana and rotating it about an axis along $\k_F$ at the polar nodes).  We will call them nodal ``lentils'', in keeping with the fruits \& vegetable  terminology.  For non-spherical Fermi surfaces, lentils will be distorted from shapes given here, but topology will remain the same.

\begin{figure}[tb]
    \centering
 \includegraphics[width=\linewidth]{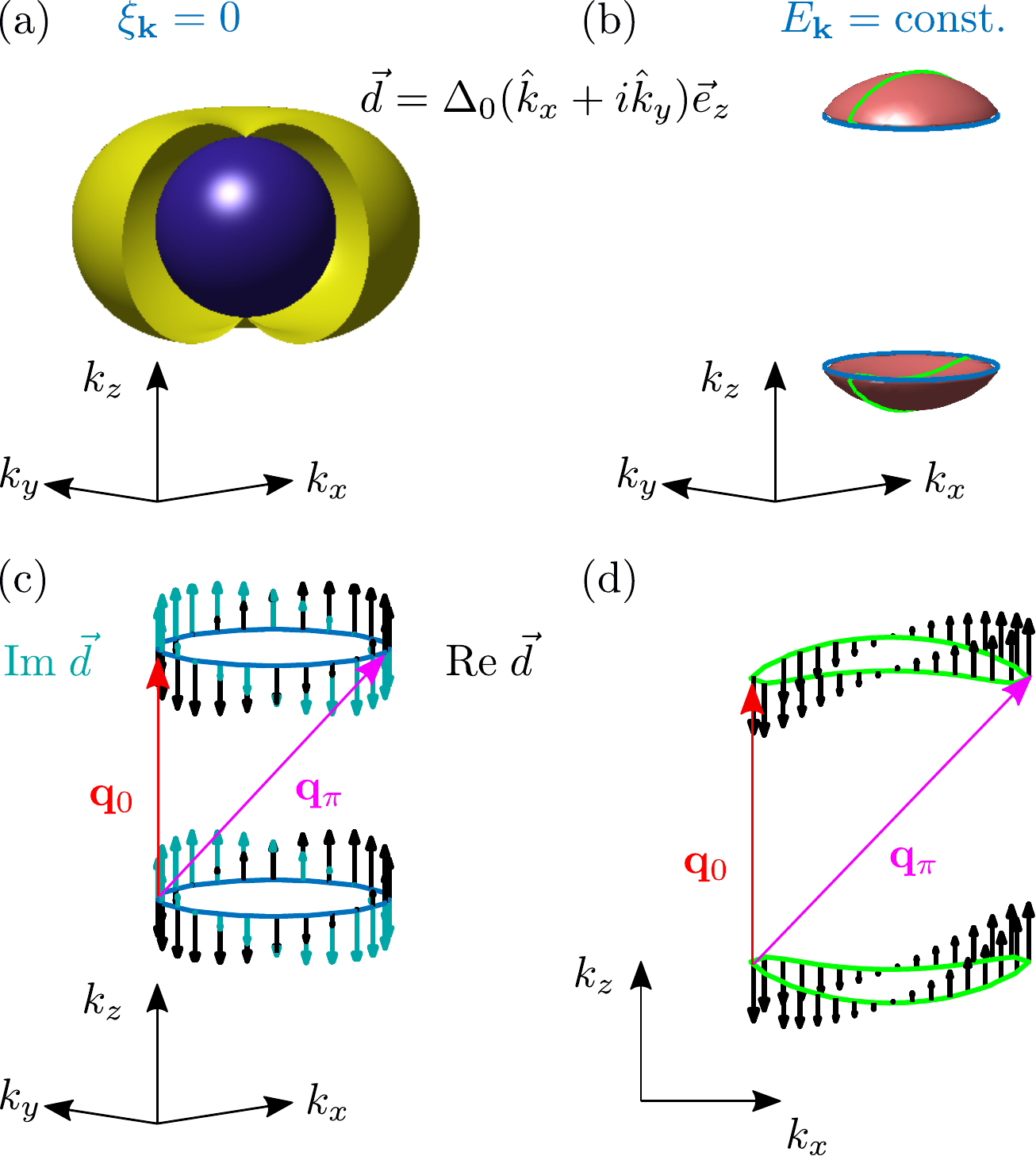}
    \caption{QPI from the ABM state (a) Fermi surface (blue) and magnitude of the $d$-vector in the ABM state with $\vec d_{\k}=\Delta_0(\hat k_x+i\hat k_y) {\vec e_z}$. (b) Contours of constant quasiparticle energies $E_{\mathbf k}$ close to zero energy are of ``lentils'' shape with two lines in the $x-y$ plane that yield dominant contributions to the density of states as in the ``octet'' model (blue line) and cut in the $x-z$ plane (green line). (c) For the dominant scattering processes, there are $q$-vectors with zero angle between $\vec d_{\mathbf k_F}$ and $\vec d_{\mathbf k_F+\q_0}^*$  and $q$-vectors where $\mathbf k_F=-\mathbf k_F+\q_\pi$ such that the vectors in the scalar product are anti-parallel (c,d).}
    \label{fig:ABM}
\end{figure}

The edge of the CCE is for $\k$ directly on the Fermi surface, i.e. $\xi_{\k}=0$, a circle of radius $k_1(\omega)=k_F\omega/\Delta_0$, one at each pole, see Fig.~\ref{fig:ABM}(b). This line plays the role of the banana tips in the two-dimensional model and dominates the scattering in the QPI.

The intra-CCE scattering processes $\q_i$ in principle connect points on the CCE with other points on the same CCE. Due to the axial symmetry, this must lead to a small ring of intensity in $\delta\rho(\q)$ centered at ${q_x}={q_y}=0$. Using $\vec d_{\k}=\Delta_0 \sin \vartheta {\vec e_z}$ and $k_1=\sin\vartheta k_F$, we find the radius of the  CCE and therefore the radius in the QPI intensity as $2k_1=2k_F \omega/\Delta_0$.
For fixed $\q$, we can examine the dot product in the density of states modulations Eq. (\ref{eq_delta_rho})
\begin{equation}
    {\vec d}_{\k}\cdot {\vec d}_{\k+\q}^* =
    \Delta_0^2 (k_x+ik_y)(k_x+q_x -ik_y -i q_y) 
\end{equation}
One sees that the product has a term whose phase varies as $\k$ goes around the circle. For $\q_0$ ($\q_\pi$) as shown in Fig. \ref{fig:ABM}, the $\vec d$-vectors of initial and final state are parallel (anti-parallel)
\begin{eqnarray}
   {\vec d}_{\k} &=& -i\Delta_0 k_1 \vec e_z\nonumber \\
   {\vec d}_{\k+\q_{0/\pi}} &=&
   \pm i\Delta_0 k_1 \vec e_z.
\end{eqnarray}
Thus, the dot product is (positive) negative  ${\vec d}_{\k}\cdot {\vec d}_{\k+\q_{0/\pi}}^* =\pm \Delta_0^2 k_1^2 $.
On the other hand, for other values of $\k$, the complex dot product will acquire a phase, and it is clear that the total contribution to $\rho(\q,\omega)$ from integrating $\k$ around the ring $\int_{\k\in{ \circ}} {\vec d}_{\k}\cdot {\vec d}_{\k+\q}^*  $ will give just the {\it positive} constant $\Delta_0^2 k_1^2$ for any $\q$.  So this intra-CCE scattering should produce a circle in $\q$-space of radius $2k_1$ which is expected to be an odd function of $\omega$ as discussed in the original work on QPI in singlet superconductors \cite{Hirschfeld2015}, but information about the relative orientation of the $\vec d$-vectors is averaged out. This is essentiallly due to the larger phase space in the three dimensional system where the energy-momentum conservation still allows for summing scattering processes with different directions of the $\vec d$-vector.

We have focused on processes across the Fermi surface, i.e. scattering with finite $q_z$.
Since the ABM $d$-vector is independent of $k_z$, also processes within the same lentil ($q_z=0$) are subject to exactly the same discussion as the ones across the Fermi surface. Thus, the expectation here is that there will be circles of intensity in $\delta\rho(\q,\omega)$ at $q_z=\pm 2k_z$ and $q_z=0$, where $k_z^2=k_F^2 - k_1^2$.

\subsubsection{Planar state}
There were two special things about the ABM state: First, $\vec d$ pointed only in one direction and second, it was independent of $k_z$. Here, we investigate how the relaxation of first condition affects things by examining one of the planar states,
\begin{equation}
    \vec d_{\k} = \Delta_0 ({\vec e_x}k_x +{\vec e_y} k_y).
\end{equation}
Note this state is real and has exactly the same magnitude, $|\vec d_{\k}^\mathrm{ABM}|^2=|\vec d_{\k}^\mathrm{planar}|^2$ and the same point nodes at the poles. The $\vec d$-vector points radially around any line of latitude on the sphere, in particular the edges of our lentils.  Again we construct \begin{equation}
     {\vec d}_{\k}\cdot {\vec d}_{\k+\q}^* = \Delta_0^2(k_1^2 + k_xq_x + k_y q_y),
\end{equation}
which will clearly average to $\Delta_0k_1^2$ as before. So, the winding of the $\vec d$-vector around the nodes makes no difference.

\subsubsection{$A_u$ and $B_{iu}$ order parameters with spherical Fermi surface}
\begin{figure}[tb]
    \centering
 \includegraphics[width=\linewidth]{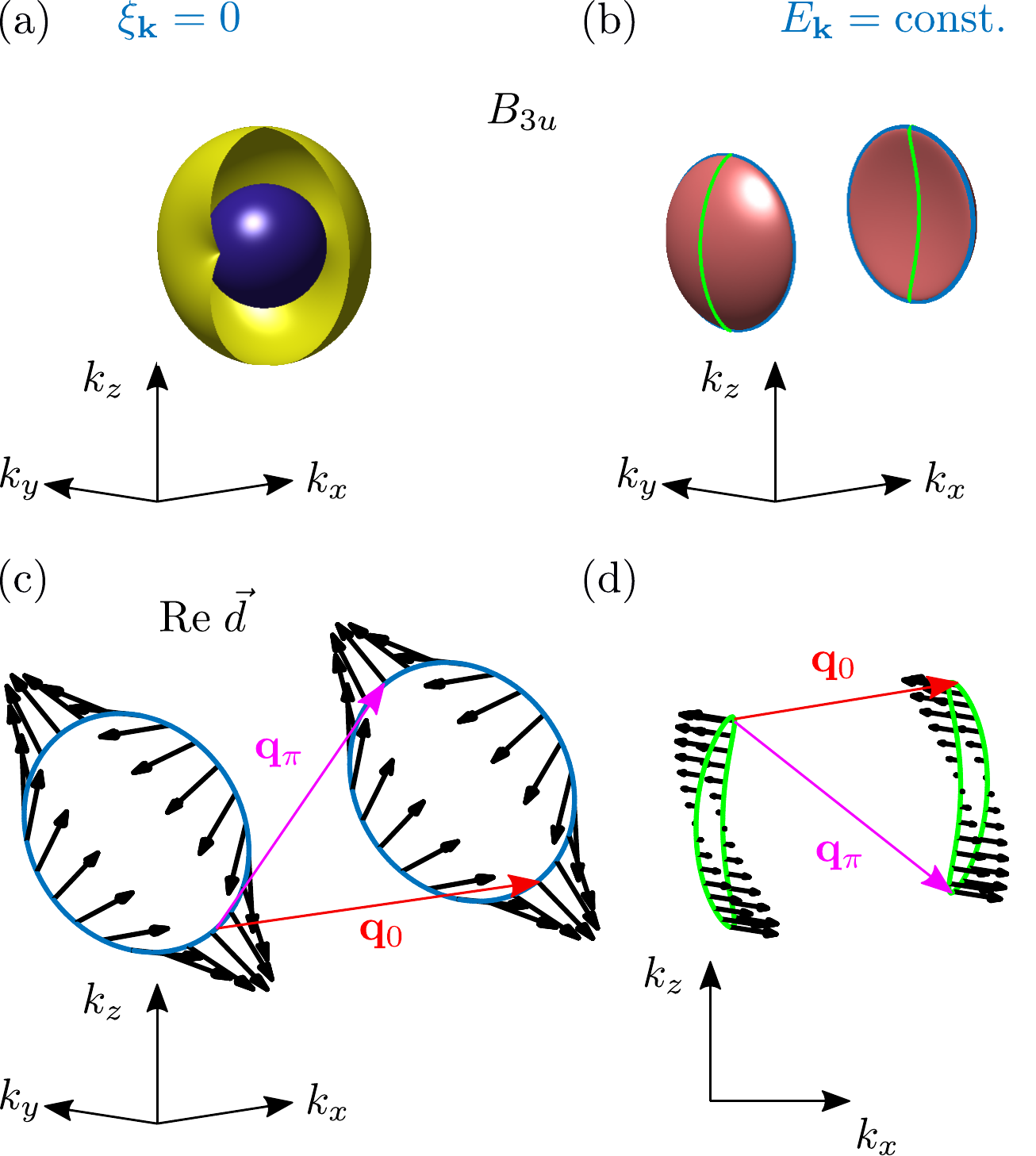}
    \caption{QPI in B$_{3u}$ state (a) Fermi surface and gap magnitude of a B$_{3u}$ state with $\vec d_{\k}=(p_1k_x k_y k_z, p_2k_z, p_3 k_y) $. (b) Contours of constant quasiparticle energies $E_{\mathbf k}$ close to zero energy are of ''lentil`` shape with two lines in the $y-z$ plane that yield dominant contributions to the density of states as in the ''octet`` model (blue line) and cut in the $x-z$ plane (green line). (c) For the dominant scattering processes, there are $q$-vectors with zero angle between $\vec d_{\mathbf k_F}$ and $\vec d_{\mathbf k_F+\q_0}^*$  and $q$-vectors where $\mathbf k_F=-\mathbf k_F+\q_\pi$ such that the vectors in the scalar product are anti-parallel (c,d).}
    \label{fig:B3u_1}
\end{figure}

\begin{figure}[tb]
    \centering
 \includegraphics[width=\linewidth]{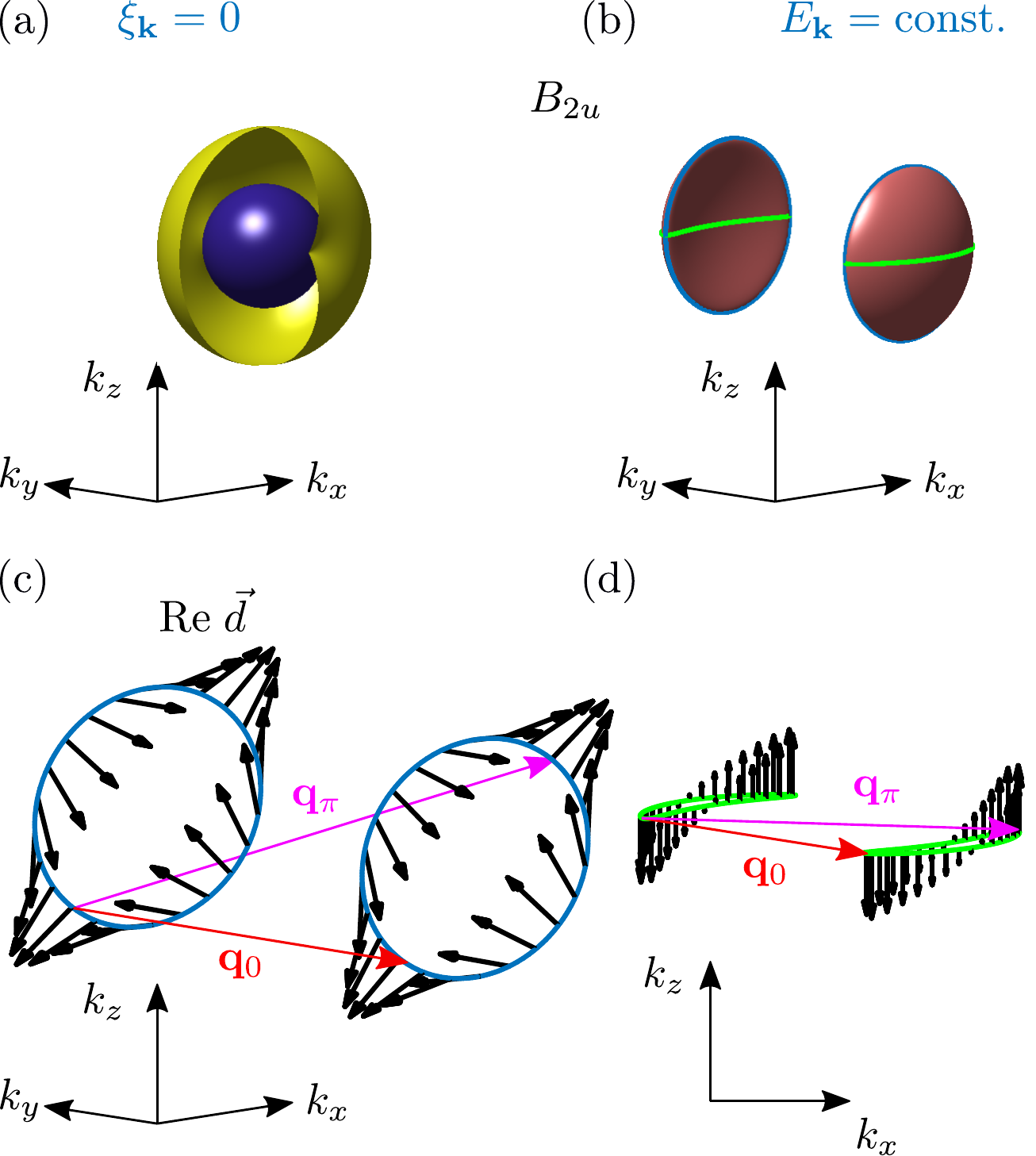}
    \caption{QPI in B$_{2u}$ state (a) Fermi surface and gap magnitude of a B$_{2u}$ state with $\vec d_{\k}=(p_1k_z, p_2k_x k_y k_z, p_3 k_x) $. (b) Contours of constant quasiparticle energies $E_{\mathbf k}$ close to zero energy are of ''lentil`` shape with two lines in the $x-z$ plane that yield dominant contributions to the density of states as in the ''octet`` model (blue line) and cut in the $x-y$ plane (green line). (c) For the dominant scattering processes, there are $q$-vectors with zero angle between $\vec d_{\mathbf k_F}$ and $\vec d_{\mathbf k_F+\q_0}^*$  and $q$-vectors where $\mathbf k_F=-\mathbf k_F+\q_\pi$ such that the vectors in the scalar product are anti-parallel (c,d).}
    \label{fig:B2u_1}
\end{figure}

\begin{figure}[tb]
    \centering
 \includegraphics[width=\linewidth]{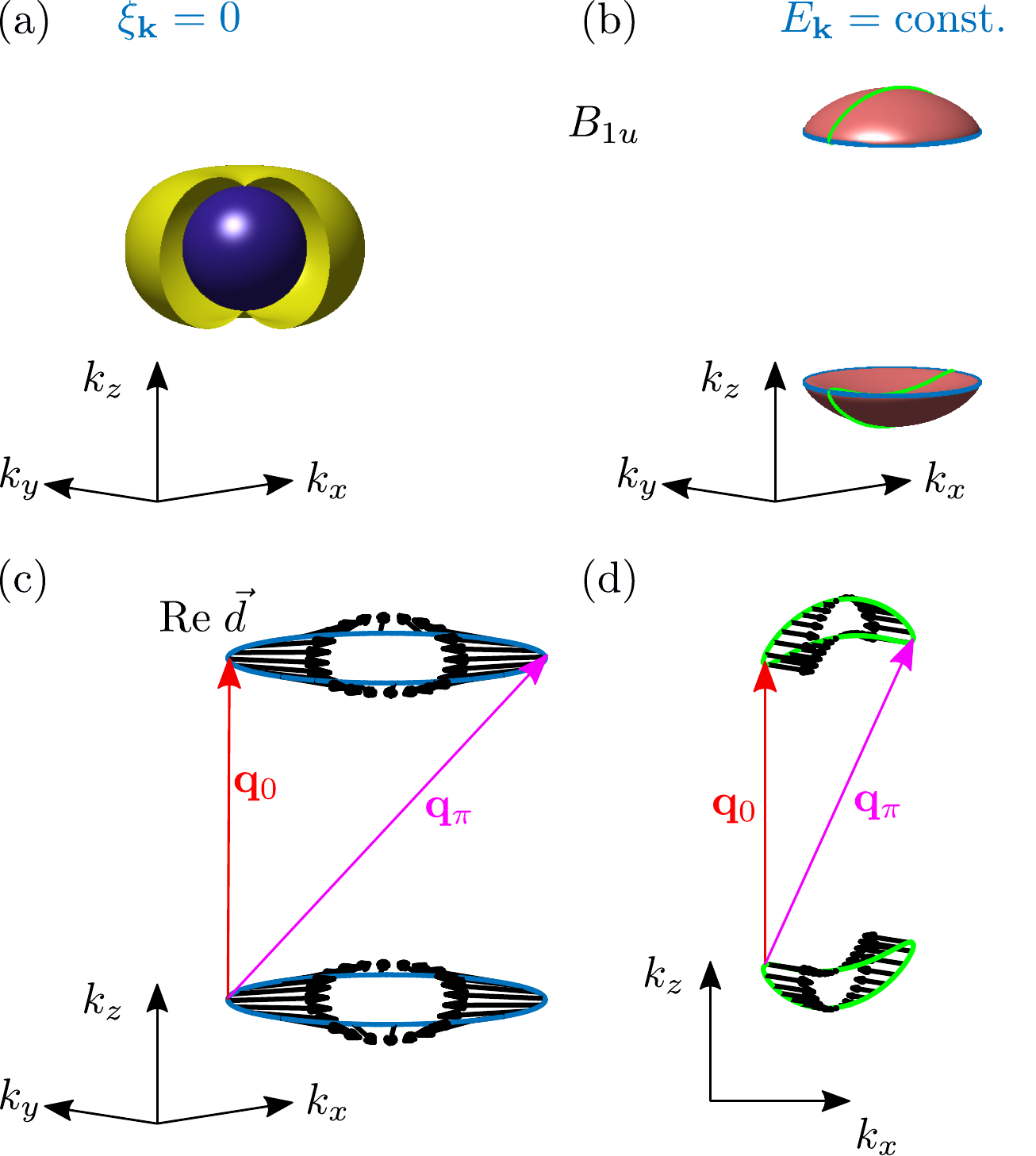}
    \caption{QPI in B$_{1u}$ state (a) Fermi surface and gap magnitude of a B$_{1u}$ state with $\vec d_{\k}=(p_1k_y, p_2k_x, p_3 k_x k_y k_z) $. (b) Contours of constant quasiparticle energies $E_{\mathbf k}$ close to zero energy are of ''lentil`` shape with two lines in the $x-y$ plane that yield dominant contributions to the density of states as in the ''octet`` model (blue line) and cut in the $x-z$ plane (green line). (c) For the dominant scattering processes, there are $q$-vectors with zero angle between $\vec d_{\mathbf k_F}$ and $\vec d_{\mathbf k_F+\q_0}^*$  and $q$-vectors where $\mathbf k_F=-\mathbf k_F+\q_\pi$ such that the vectors in the scalar product are anti-parallel (c,d).}
    \label{fig:B1u_1}
\end{figure}

\begin{figure}[tb]
    \centering
 \includegraphics[width=\linewidth]{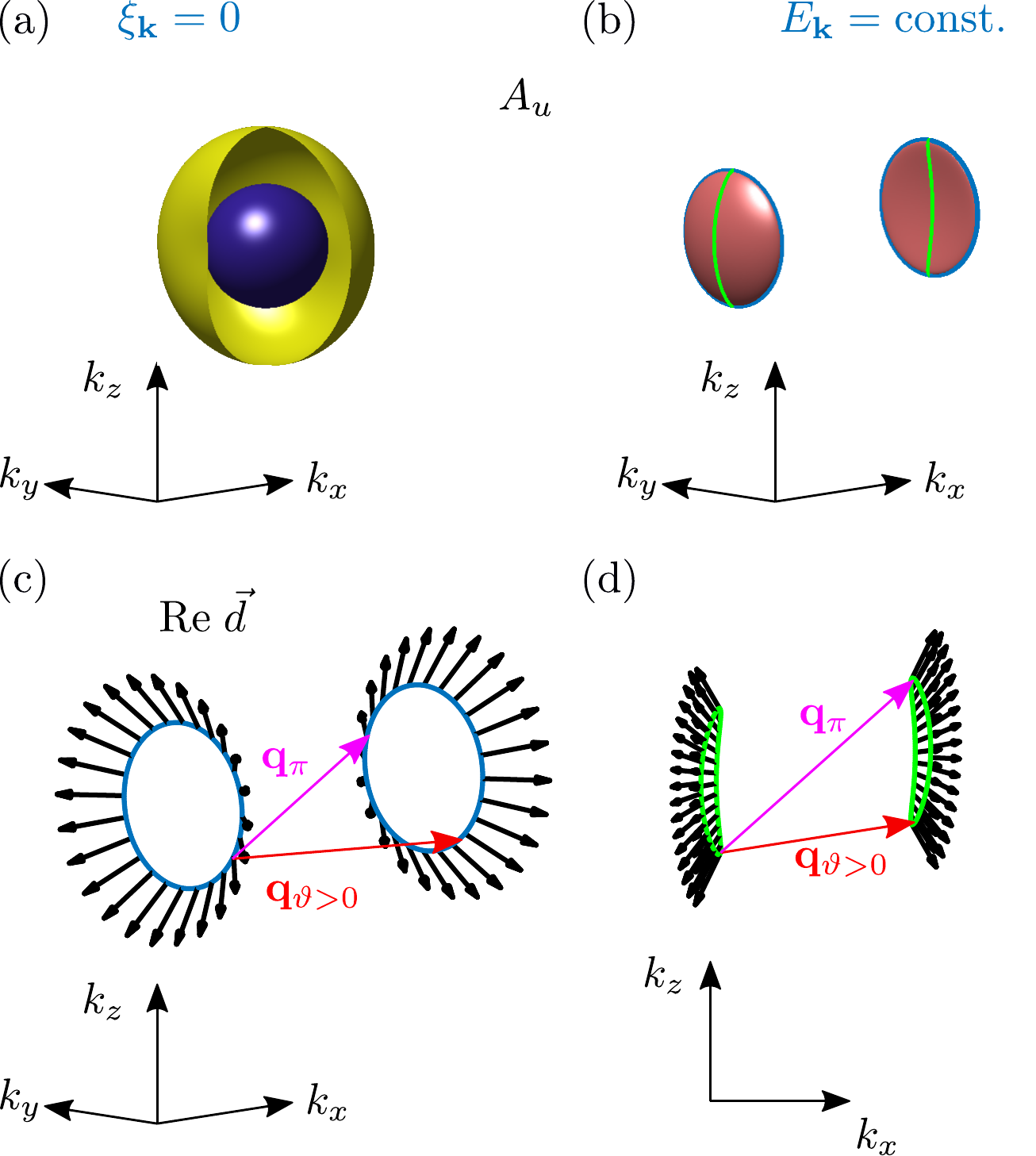}
    \caption{QPI in A$_{u}$ state (a) Fermi surface and gap magnitude of a A$_{u}$ state with $\vec d_{\k}=(p_1k_x , p_2k_y, p_3 k_z) $.  The parameters $p_1,p_2,p_3$ are chosen to give a gap minimum along the $k_x$ axis.  (b) Contours of constant quasiparticle energies $E_{\mathbf k}$ slightly above this minimum are of ''lentil`` shape with two lines in the $y-z$ plane that yield dominant contributions to the density of states as in the ''octet`` model (blue line) and cut in the $x-z$ plane (green line). (c) For the dominant scattering processes, there are only $q$-vectors with finite angle between $\vec d_{\mathbf k_F}$ and $\vec d_{\mathbf k_F+\q_\theta}^*$, the largest angle is achieved for $q$-vectors where $\mathbf k_F=-\mathbf k_F+\q_\pi$ such that the vectors in the scalar product are anti-parallel (c,d).}
    \label{fig:Au_1}
\end{figure}

\begin{table}
\begin{tabular}{|c|c|c|}
\hline
$\Gamma $&  Gap function $\vec d_{\k}$  & $\text { Nodes }$ \\
\hline $A_{1 u}$ & $\left(p_1 k_x, p_2 k_y, p_3 k_z\right)$ & $\text { Accidental }$ \\
\hline $B_{1 u}$ &$ \left(p_1 k_y, p_2 k_x, p_3 k_x k_y k_z\right)$ & $z$ $\text {-axis }$ \\
\hline $B_{2 u}$ &$ \left(p_1 k_z, p_2 k_x k_y k_z, p_3 k_x\right)$ & $y \text {-axis } $\\
\hline$ B_{3 u}$ & $\left(p_1 k_x k_y k_z, p_2 k_z, p_3 k_y\right)$ & $x \text {-axis } $\\
\hline
\end{tabular}
\caption{Allowed low-order basis functions for orthorhombic symmetry in the strong spin-orbit coupling limit\cite{hillier2012nonunitary} on sphere as discussed in view of the small energy QPI. For order parameters on a lattice, the powers of $k_x$, $k_y$, $k_z$ have to be replaced by lattice harmonics that transform like $X$, $Y$ and $Z$.}
\label{tab:dvector}
\end{table}
For simplicity, we continue the discussion on the example of a $B_{2u}$ order parameter with a quadratic (normal state) dispersion. The conclusions will, however, be unchanged for arbitrary dispersions as long as one stays at small energies and expands around the nodes which in our case for a node at ${\k}_\mathrm{node}=k_F(0,1,0)$ yields
\begin{eqnarray}
    E_{\k}^2 &=& v_F^2(k_\perp-k_\mathrm{node})^2 + |\vec d^2| \nonumber \\
    &=& v_F^2\delta k_y^2 + \Delta_0^2 (p_1^2\delta k_z^2+p_2^2\delta k_x^2 k_F^2\delta k_z^2+p_3^2\delta k_x^2)\nonumber\\
    &\simeq& v_F^2\delta k_y^2 + \Delta_0^2 (p_1^2\delta k_z^2+ p_3^2\delta k_x^2).
\end{eqnarray}
We have neglected $d_y(\mathbf k)$ since close to the node, $k_x$ and $k_z$ are both small, thus $d_y$ is quadratic in this small quantity. Within this approximation, the contour of constant energy is again of the form of a lentil, squashed by the factor $p_1/p_3$ in one direction, and we are getting a planar-like texture of the $\vec d$-vector as shown in Fig. \ref{fig:B2u_1}. The circles of high DOS are evidently replaced by ellipses (blue lines in Figs. \ref{fig:B3u_1}, \ref{fig:B2u_1}, \ref{fig:B1u_1}). Again, we construct the scalar product and neglect terms that are quadratically small close to the nodal axis,
\begin{align}
     {\vec d}_{\k}\cdot {\vec d}_{\k+\q}^* &
     \simeq  \left(k_1^2 +p_1^2 k_zq_z +p_3^2k_x q_x\right),
\end{align}
where now $k_1^2=p_1^2k_z^2+p_3^2 k_x^2=\omega^2$.  So it appears that the extra terms will also average to zero when one integrates over $k_x,k_z$. The other order parameters $B_{1u}$ and $B_{3u}$ are just rotated versions of the present case with rescaling of the lentils and ellipses according to the value of $p_i$. An exception is the $A_u$ order parameter which is fully gapped, so the contours of constant energy can only be considered for finite $\omega$. In this case $\vec d$-vector has no planar texture on the ellipse of high DOS, see Fig. \ref{fig:Au_1}, so the analysis is more complicated and depends on the relative size of the $p_i$ which we will not follow up here. Similar complications also occur once $\omega$ is large enough that the texture of the $\vec d$-vector is not planar any more for the $B_{iu}$ states.

\section{Model for the normal and superconducting states of UT\lowercase{e}$_2$}

\subsection{Normal State}

Our starting point is the tight-binding model \cite{Theuss_2024,Christiansen}
\begin{align}
    H_N(\mathbf k) = \begin{pmatrix}
        H_{\mathrm{U}} && H_{\mathrm{Te}-\mathrm{U}} \\
        H_{\mathrm{Te}-\mathrm{U}}^\dag && H_{\mathrm{Te}}
    \end{pmatrix} \label{eq:H_normal},\
\end{align}
where $H_{\rm U}$ and $H_{\rm Te}$ describe the uranium and tellurium bands close to the Fermi surface and are given by 
\begin{align}
    H_{\mathrm{U}} =\phantom{-} \rho_0\;&[\mu_\mathrm U - 2t_\mathrm U \cos k_x - 2t_\mathrm {ch,U}\cos k_y] \notag \\
    -\rho_1\;&[ \Delta_\mathrm U + 2t_U' \cos k_x + 2t_\mathrm {ch,U}' \cos k_y \notag \\
    &\;- 4t_\mathrm {z,U} \cos(k_x/2)\cos(k_y/2)\cos (k_z/2)] \notag \\
    - \rho_2\;& 4 t_\mathrm {z,U}\cos(k_x/2)\cos(k_y/2)\sin(k_z/2) \notag    \\
    \equiv\phantom{-}\rho_0 \;&f_{\mathrm{U},0}(\mathbf{k}) + \rho_1\;f_{\mathrm{U},1}(\mathbf{k}) +\rho_2\;  f_{\mathrm{U},2}(\mathbf{k})\label{eq:HU},\
\end{align}
and
\begin{align}
    H_{\mathrm{Te}} =\phantom{-} \rho_0\; &(\mu_{\mathrm{Te}} - 2t_{ch,\mathrm{Te}}\cos k_x) \notag \\
    -\rho_1  \; & [\Delta_{\mathrm{Te}} - t_{\mathrm{Te}}\cos k_y  \notag\\
    &-2t_{z,\mathrm{Te}}\cos(k_x/2)\cos(k_y/2)\cos(k_z/2)]\notag \\
    -\rho_2\; & t_{\mathrm{Te}}\sin k_y  \notag\\
    \equiv\phantom{-}\rho_0 \;& f_{\mathrm{Te},0}(\mathbf{k}) + \rho_1 \;f_{\mathrm{Te},1}(\mathbf{k})+\rho_2\; f_{\mathrm{Te},2}(\mathbf{k}) \label{eq:HTe}.\
\end{align}

The hybridization between the uranium and tellurium sector is given by $H_{\mathrm{Te}-\mathrm{U}}=\delta (\rho_0+\rho_1)$, chosen such that the $\mathcal M_z$ and $\mathcal M_y$ mirror symmetries are preserved, as explained in \cite{Christiansen}. All parameters are written in Tab.~\ref{tab:ute2parameters}. 

\begin{table*}
    \centering
    \begin{tabular}{|c|c|c|c|c|c|c|c|c|c|c|c|c|} \hline
         $\mu_U$ & $\Delta_U$ & $t_U$ & $t_U'$ & $t_{ch,U}$ & $t_{ch,U}'$ & $t_{z,U}$ & $\mu_{\mathrm{Te}}$ & $\Delta_{\mathrm{Te}}$ & $t_{\mathrm{Te}}$ & $t_{ch,\mathrm{Te}}$ & $t_{z,\mathrm{Te}}$ & $\delta$  \\ \hline
         0.4 & 0.35 & 0.15 & 0.08 & 0.01 & 0 & 0.08 & -1.8 & -1.5 & -1.5 & 0 & -0.05 & 0.045
     \\ \hline\end{tabular}
    \caption{Parameters (in units of $\rm eV$) for the normal-state tight binding model}
    \label{tab:ute2parameters}
\end{table*}

\subsection{Superconducting Phases}

The superconducting phase in UTe$_2$ is described by the BdG Hamiltonian as written in Eq. (\ref{eq:bdgham}), generalized to the multi-band case. Restricting to spin-triplet orders, $\Delta(\mathbf{k})$ is written as in Eq. (\ref{eq_delta}) where $\vec d_{\mathbf{k}}$ is now a three-vector with each element a matrix in sublattice and U-Te space. By considering the representation of the D$_{2h}$ point group on the sublattice and U-Te space, one can construct all possible combinations of sublattice and momentum functions such that the product transforms like $X$, $Y$, $Z$ or $XYZ$, obtaining the result in Tab.~\ref{tab:basisfunctions} \cite{Christiansen}. By combining with the appropriate spin Pauli matrix according to Tab.~\ref{tab:dvector}, one can then write all symmetry-allowed terms for each superconducting phase. Besides the symmetry-imposed nodes, the shape of the Fermi surface of this material implies additional nodes. As was explained in Ref.~\cite{Christiansen}, the location of the additional nodes is not fixed by symmetry, i.e. the additional nodes move depending on the coefficients of the superconducting terms. In Fig.~\ref{fig:additionalnodes}, we show the location of the additional nodes for different choices of $d_{z,\mathrm{Te}},d_{z,\mathrm{U}}  \sim (1-\alpha)f_{x,1}+\alpha f_{x,2}$ with $\alpha=0,0.5,1$, where $f_{x,1}(\mathbf{k}) = \sin(k_x/2)\cos(k_y/2)\cos(k_z/2)$ and $f_{x,2}(\mathbf{k}) = \sin(k_x)$. Different $\alpha$ correspond to different ratios of nearest- to next-nearest neighbor pairing bonds.

\begin{table*}[tb]
    \begin{tabular}{|l|l|l|l|}
        \hline
                 &   U sector&Te sector &U-Te sector\\
                 \hline
                 $X$& 
             $f_x \rho_{1/0}$, $f_{xz}\rho_2$&$f_x\rho_{1/0}$, $f_{xy} \rho_2$&$f_x(\rho_0+\rho_1)$, $f_{xyz}(\rho_0-\rho_1)$\\ \hline
         $Y$& $f_y\rho_{1/0}$, $f_{yz}\rho_2$&$f_y\rho_{1/0}$, $f_1\rho_2$&$f_y(\rho_0+\rho_1)$, $f_z(\rho_0-\rho_1)$\\ \hline
         $Z$& $f_z \rho_{1/0}$, $f_1\rho_2$&$f_z\rho_{1/0} $, $f_{yz}\rho_2$&$f_z(\rho_0+\rho_1)$, $f_{y}(\rho_0-\rho_1)$\\
         \hline
         $XYZ$ & $f_{xyz}\rho_{1/0}$, $f_{xy}\rho_2$&$f_{xyz}\rho_{1/0}$, $f_{xz}\rho_2$&$f_{xyz}(\rho_0+\rho_1)$, $f_x(\rho_0-\rho_1)$\\
         \hline 
         
     \end{tabular}

     \caption{Combinations of sublattice Pauli matrices $\rho_i$ and momentum functions such that the combination transforms as either $X$, $Y$, $Z$ or $XYZ$, where e.g. $X$ indicates that it transform like $k_x$. Similarly, $f_x$ indicates a momentum function that transform like $k_x$ which we choose either as $f_{x,1}(\mathbf{k}) = \sin(k_x/2)\cos(k_y/2)\cos(k_z/2)$ or $f_{x,2}(\mathbf{k}) = \sin(k_x)$. $f_1$ indicates a function that transforms trivially under all point group transformations.}
     
    \label{tab:basisfunctions}
\end{table*}

\section{The (0-11) Surface}

\subsection{Lattice Transformation}

We transform to the primitive unit cell defined by the lattice vectors
\begin{equation}
    \begin{pmatrix}
        \Vec{r}_1 & \Vec{r}_2 & \Vec{r}_3
    \end{pmatrix}
    = \begin{pmatrix}
        -a/2 & a/2 & a/2 \\
        b/2 & -b/2 & b/2 \\
        c/2 & c/2 & -c/2
    \end{pmatrix},
\end{equation}
written in the basis of the Cartesian unit vectors $\{\hat e_x, \hat e_y, \hat e_z\}$ and $(a,b,c)$ are the lattice constants along the three crystal axes. The corresponding reciprocal lattice vectors are
\begin{equation}
    \begin{pmatrix}
        \Vec{m}_1 & \Vec{m}_2 & \Vec{m}_3 
    \end{pmatrix}
    = 
    \begin{pmatrix}
        0 & 1/a & 1/a \\
        1/b & 0 & 1/b \\
        1/c & 1/c & 0
    \end{pmatrix} ,\
\end{equation}
which implies the following coordinate transformations 
\begin{equation}
    \begin{pmatrix}
        k_x \\ k_y \\ k_z
    \end{pmatrix}
    =
    \begin{pmatrix}
        k_2 + k_3 \\
        k_3 + k_1 \\
        k_1 + k_2
    \end{pmatrix}.
\end{equation}
Inserting this transformation means that the BdG Hamiltonian is written as a function of the Fourier coefficients $\mathrm{e}^{\pm i k_i}$ for $i=1,2,3$. The corresponding real space model is found by the replacement $\mathrm{e}^{-ik_i}\mapsto T_{\vec r_i}$ and $\mathrm{e}^{ik_i}\mapsto T_{\vec r_i}^T$, where the translation operators, $T_{\vec {r_i}}$ act on the real-space basis vectors, $\ket{\vec r}$, as 
\begin{equation}
    T_{\vec r_i}\ket{\mathbf{r}} = \ket{\mathbf r + \vec r_i}~,
\end{equation}
where $\mathbf{r} = \sum_i \vec r_i n_i$, with $\vec r_i$ the lattice vectors defining the unit cell and, $n_i$, integers. We now wish to transform to a unit cell that allows us to compute the surface Green's function. To do so, we choose two of the lattice vectors to lie in the surface plane, i.e. orthogonal to the normal vector while the last one is out of plane. We choose the in-plane vectors as $\vec r_1$ and $\vec r_+ = \vec r_2 + \vec r_3$ and the out-of-plane vector as $\vec r_3$. Using that $T_{\vec r_2} = T_{\vec a }T^T_{\vec r_3}$ allows us to write the BdG Hamiltonian as 
\begin{equation}\label{eq:hamiltontridiag}
    H(\mathbf k^{\parallel}) = H_0(\mathbf k^{\parallel}) + V(\mathbf{k^\parallel})T_{\vec r_3} + V^\dag(\mathbf{k^\parallel})T_{\vec r_3}^T ,\
\end{equation}
written in terms of the basis vectors 
\begin{equation}
    \ket{\mathbf k^{\parallel},n_3} = \sum_{n_1,n_+}\mathrm{e}^{i\mathbf{k}^{\parallel}\cdot \mathbf{r}^{\parallel}}\ket{\mathbf{r}},
\end{equation}
with $\mathbf{r}^{\parallel} = n_1 \vec r_1 + n_+ \vec r_+ $ and $\mathbf{k}^\parallel=k_1 \vec m_1 + k_+ \vec m_+$, where the reciprocal lattice vectors can be chosen as 
\begin{equation}
    \mqty(\vec m_1 && \vec m_+ ) = \mqty( 0 && 1/a \\ 1/b && 1/(2b) \\ 1/c && 1/(2c)). 
\end{equation}
Finally, we perform the transformation 
\begin{equation}\label{eq:starcoords}
    \mqty(\vec m_{c^*} && \vec m_{a^*}) = \mqty(\vec m_1 && \vec m_+ - \vec m_1 /2 ) ,\  
\end{equation}
so the reciprocal vectors are orthogonal. The corresponding coordinates transform as 
\begin{equation}
    \mqty( k_{c^*} \\ k_{a^*}) = \mqty( k_1 + \frac{k_+}{2} \\ k_+ ) .\
\end{equation}
The real-space vectors corresponding to the reciprocal vectors in Eq.~\eqref{eq:starcoords}, that lie in the surface plane are $\vec r_{c^*} = (0,b/2,c/2)^T$ and $\vec r_{a} = (a,0,0)$, so we identify $k_{a^*}=k_x$. 

The Green's function corresponding the Hamiltonian in Eq.~\eqref{eq:hamiltontridiag} is written as
\begin{align}\label{eq:Gtridiag}
    G(\omega,\mathbf k^{\parallel})  &=
    \frac{1}{\omega + i\eta - H(\mathbf k^{\parallel})}  \notag \\ 
    &=\begin{pmatrix}
    G_s & \hdots  &  &  &  \\
    \vdots & \ddots &  &  &   \\
     & & G_{b} &  &   \\
     & & & \ddots &  \vdots \\
    & & & \hdots & \tilde{G_{s}} 
    \end{pmatrix}  ,\
\end{align}
where each block is written in the sector of constant $n_3$ such that $G_s$ ($G_b$) [$\tilde{G}_s$] corresponds to $n_3=0$ ($n_3=N_3/2$) [$n_3=N_3$] where $N_3$ is the number of layers along the $\vec r_3$ direction. Using the method described in Ref. \cite{sancho_highly_1985}, allows for efficient calculation of these three blocks of the full Green's function in the limit $N_3\to \infty$, which we will refer to as the surface, bulk, and dual surface Green's functions, respectively.
\begin{figure}[tb]
    \centering
    \includegraphics[width=\linewidth]{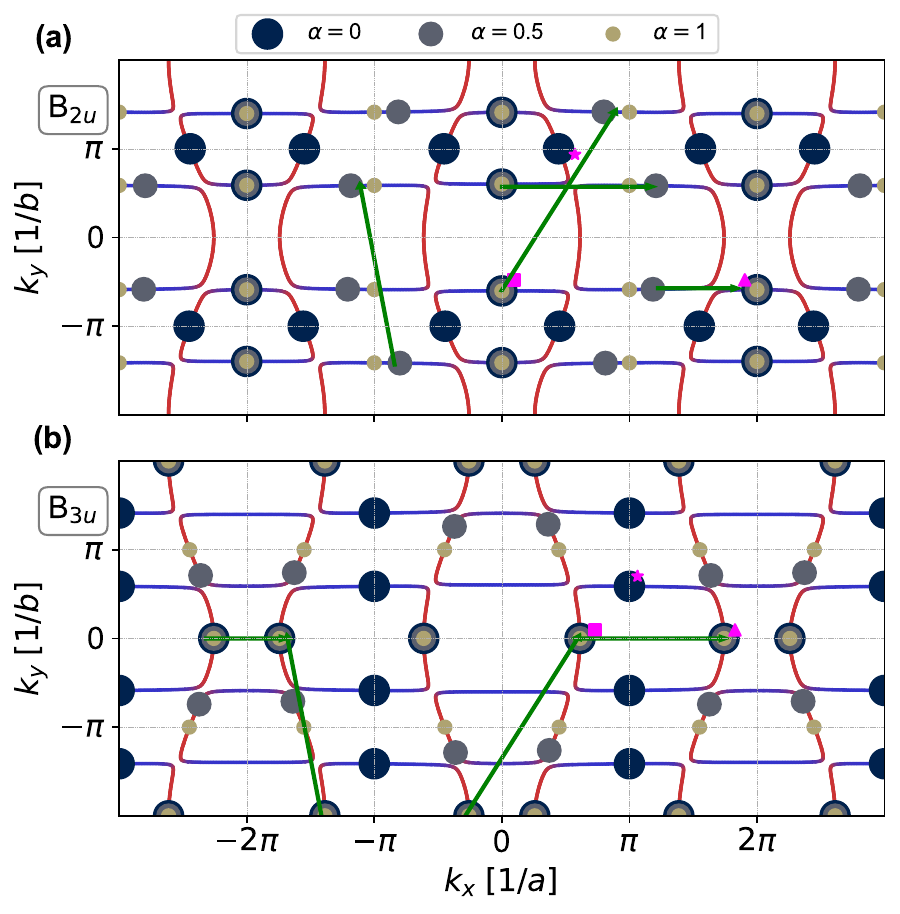}
    \caption{Uranium (red) and tellurium (blue) bands in the $k_z=0$ plane. The points indicate the nodal points of the quasiparticle spectrum for different choices of the superconducting order parameters. In the main text we use the configuration $\alpha=0$. Note that in the B$_{3u}$ phase, the $\mathbf q_i$ vectors appear from scattering between symmetry imposed nodes while for the B$_{2u}$ phase, it is only at the fine-tuned point $\alpha=0.5$, where the $\mathbf q_i$ vectors approximately appear.}
    \label{fig:additionalnodes}
\end{figure}

\subsection{QPI on the (0-11) surface}
\begin{figure*}
    \centering
    \includegraphics[width=\linewidth]{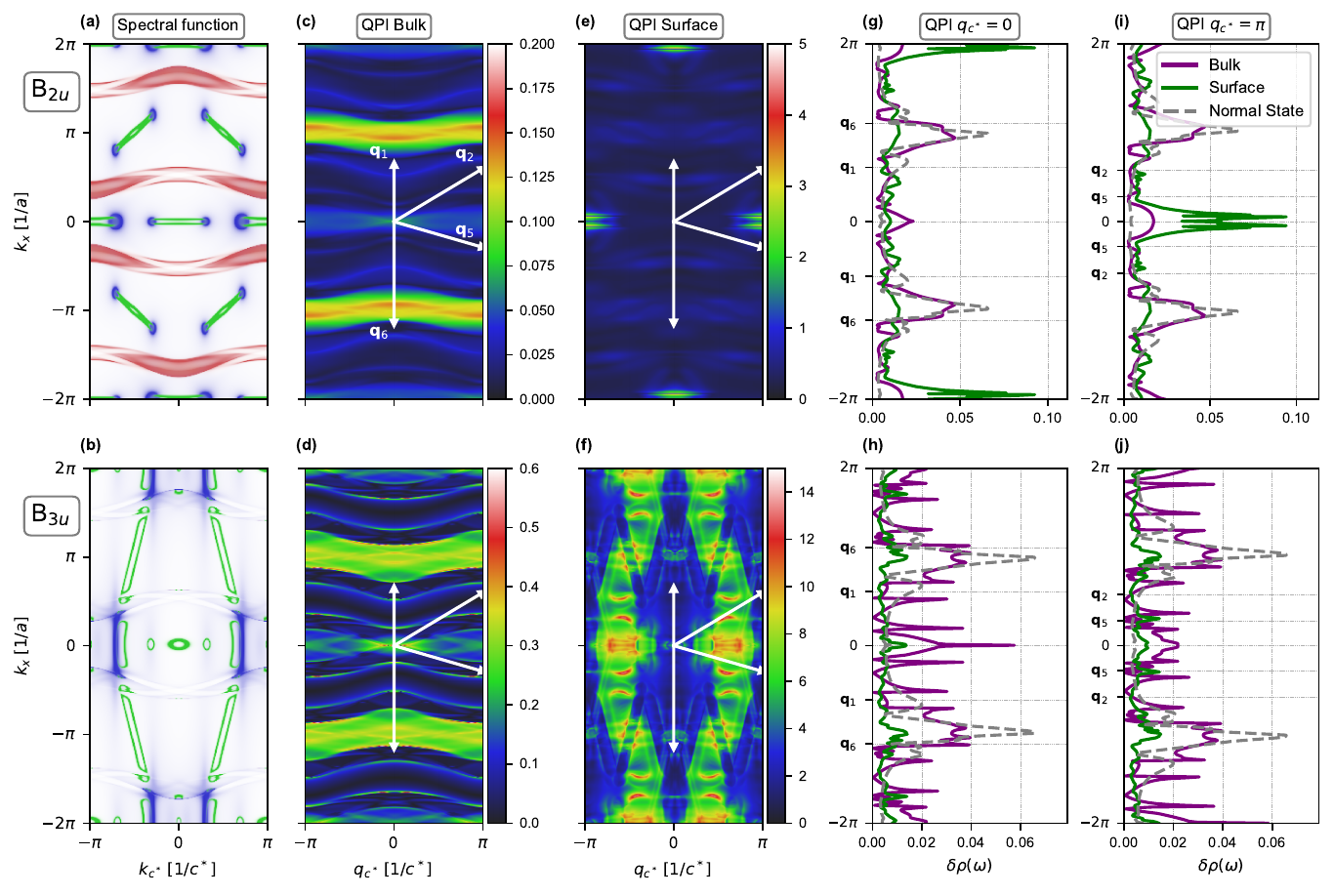}
    \caption{QPI signal and spectral functions on the (0-11) surface of UTe$_2$ in the B$_{2u}$ and B$_{3u}$ phases with $\alpha=0.5$. (a-b) The bulk U (red) and Te (blue) spectral functions along with the surface states (green). (c,d) [e,f] QPI versus surface momentum at $\omega=0.05\Delta_0$ including the surface-projected bulk [surface] states. Characteristic scattering vectors $\mathbf q_i$ are indicated by arrows. (g,h) [(i,j)] Momentum cuts of the respective panels (c,d) [e,f] at $q_{c^*}=0, \pi$. In panels (g-j) we have also indicated the locations of $\mathbf q_i$. As seen, even at this fine-tuned case where the additional nodes in the B$_{2u}$ phase are consistent with the $\mathbf{q}_i$ vectors, the B$_{3u}$ phase still seems to agree better with the intensity peaks seen experimentally~\cite{SeamusQPI}.}
    \label{fig:qpi_alpha05}
\end{figure*}

Having obtained the homogeneous Green's functions, we turn to QPI signals induced by impurities on the \mbox{(0-11)} surface. The presence of impurities is captured within the standard Born approximation as $G^{\rm imp} = G[1-VG]^{-1}\approx G+GVG$, where $V$ is the impurity potential. Choosing a point-like impurity located at the origin of the surface plane, we calculate the Green's function on the surface as
\begin{equation}
    G_s^{\rm imp}(\mathbf{r}_1^{\parallel},\mathbf{r}_2^{\parallel}) = G_s(\mathbf{r}_1^{\parallel}-\mathbf{r}_2^{\parallel}) + G_s(\mathbf{r}_1^{\parallel})V_0\tau_3 P_{\beta}G_s(-\mathbf{r}_2^{\parallel})\,,
\end{equation}
where $V_0$ is the strength of the impurity potential, $\tau_3=\mathrm{diag}(1,-1)$ is the third Pauli matrix in Nambu space and $P_\beta$ is the projector to one of the sectors in sublattice and U-Te space. The modulation in the LDOS due to the impurity at position $\beta$, is written as
\begin{equation}
    \delta \rho_{\beta}(\mathbf{r}^{\parallel},\omega) = -\frac{V_0}{\pi}\mathrm{Im}\:\mathrm{Tr}\left(\tau_e G(\mathbf{r}^{\parallel},\omega)\tau_3 P_{\beta}G(-\mathbf{r}^{\parallel},\omega)\right) ,\
\end{equation}
$\tau_e = (\tau_0 + \tau_3)/2$ projects to the electronic sector and $G=G_{s}$ or $G=G_b$ to obtain the surface or bulk QPI signal, respectively. In momentum space, this becomes 
\begin{equation}\label{eq:densitymod}
    \delta \rho_{\beta}(\mathbf{q}^{\parallel},\omega) = -\frac{V_0}{2\pi i} \bigl[ f_{\beta}(\mathbf q^{\parallel},\omega) - f^*_{\beta}(-\mathbf q^{\parallel},\omega)\bigr]\;,
\end{equation}
with 
\begin{equation}\label{eq:surfacefeq}
    f_{\beta}(\mathbf q^{\parallel},\omega) = \sum_{\mathbf k^\parallel} \mathrm{Tr}\qty(\tau_e G(\mathbf k^{\parallel}+\mathbf q^{\parallel},\omega)\tau_3P_\beta G(\mathbf k^{\parallel},\omega))   .\
\end{equation}
Experimentally, one would take the average over many impurities scattered uniformly over the sample. The location of the impurity would introduce an extra phase factor in Eq.~\eqref{eq:densitymod}. Averaging over all impurities can therefore be done by taking the absolute value of $\delta\rho_{\beta}$. The total QPI signal is therefore computed as 
\begin{equation}
    \delta \rho(\mathbf{q}^\parallel,\omega) = \sum_\beta \abs{ \delta\rho_{\beta}(\mathbf{q}^\parallel,\omega)} ,\
\end{equation}
where we both assume that the impurities are distributed equally over all sublattices and U/Te sectors and the density in each sector contributes equally to the signal. As long as no details about density and preferred type of impurities are available from experiments, this is the simplest assumption. 
In Fig.~\ref{fig:qpi_alpha05} we show the QPI signal as in Fig.~3 of the main text, but now with $\alpha=0.5$ so the additional nodes appear away from the high symmetry points, according to the discussion around Fig.~\ref{fig:additionalnodes}. At this fine tuned point, the additional nodes in the $\mathrm{B}_{2u}$ phase are approximately connected by the $q$-vectors. Even at this point however, only the $\mathrm{B}_{3u}$ phase features enhanced QPI peaks at the $q$-vectors, as discussed in the main text. 

\subsection{QPI signal from Surface States}

\begin{figure}
    \centering
    \includegraphics[width=\linewidth]{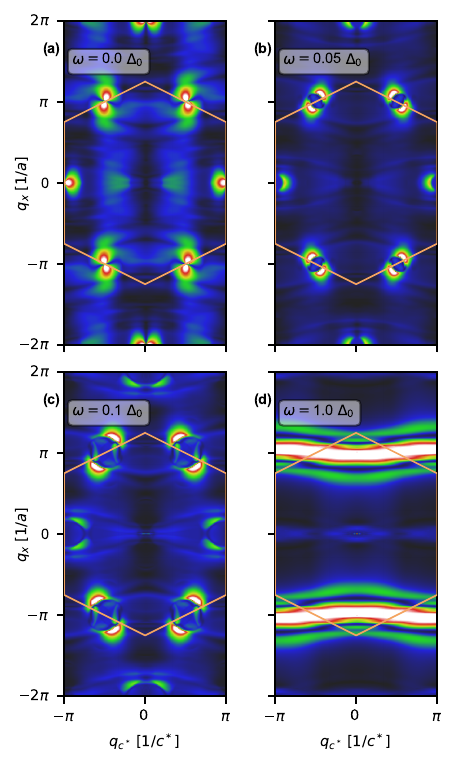}
    \caption{Surface QPI signal, $\delta\rho(\mathbf{q^\parallel,\omega})$, in the $B_{\rm 3u}$ phase at $\alpha=0$. At $\omega<\Delta_0$, the signal has distinct features related to the structure of the topological surface states.}
    \label{fig:b3usurfaceqpi}
\end{figure}

In this section, we use the low-energy expansion of the surface states close to the TRIM points to explain the main features observed in the surface QPI of the B$_{3u}$ phase as presented in Fig. 3(f) in the main text and in Fig.~\ref{fig:b3usurfaceqpi}(a-c) for a number of additional energies. These can be summarized as 
\begin{itemize}
    \item Vanishing signal at $\mathbf{q}^\parallel=(0,0)$, i.e. no scattering within the same Dirac cone.
    \item Concentrated intensity close to, but not exactly at, the TRIM point. 
    \item Intensity variation as a function of angle around the TRIM points with maxima and intensity nodes.
\end{itemize}

To do so, we exploit the time-reversal symmetry, with bulk representation $\mathcal U(\mathcal T) = \sigma_2 K$, where $K$ represents complex conjugation. 
Together with the representation of the particle-hole anti-symmetry (enforced by the BdG description) with representation $\mathcal U(\mathcal P) = K\tau_1$, the combined chiral anti-symmetry has bulk representation $\mathcal U(\mathcal P) = \tau_1\sigma_2$ \cite{Geier_2020}. At a given TRIM point, $\mathbf{k}_n^\parallel$, the BdG Hamiltonian has two zero-energy eigenstates localized at the surface \cite{Sato,Christiansen}. These can be chosen to simultaneously diagonalize $\mathcal U(\mathcal C)$, i.e. they are written as $\ket{\gamma_a^n}$ such that $\mathcal U(\mathcal C)\ket{\gamma_a^n} = a\ket{\gamma_a^n}$ for $a=\pm 1$. For any matrix $O$, such that $\mathcal U^\dag (\mathcal C)O\mathcal U(\mathcal C) = \pm O$, and for two TRIM points $\mathbf k_n^\parallel$ and $\mathbf{k}_m^\parallel$, the matrix elements of $O$ with the associated surface states obey the relation
\begin{equation}
    \mel{\gamma^n_a }{O}{\gamma^m_b} = \pm ab \mel{\gamma^n_a }{O}{\gamma^m_b} ,
\end{equation}
meaning that matrices that commute (anti-commute) with $\mathcal U (\mathcal C) $ have diagonal (off-diagonal) matrix elements only. Furthermore, by Kramer's theorem, $\mathcal U(\mathcal T)\ket{\gamma_+^n}$ is orthogonal to $\ket{\gamma_+^n}$ and should therefore be proportional to $\ket{\gamma_-^n}$, i.e. $\mathcal U(\mathcal T)\ket{\gamma_+^n}=\mathrm{e}^{i\phi}\ket{\gamma_-^n}$ and $\mathcal U(\mathcal T)\ket{\gamma_-^n}=-\mathrm{e}^{-i\theta}\ket{\gamma_+^n}$ so 
\begin{align}\label{eq:tau3melvanish}
    \mel{\gamma_-^n}{\tau_3}{\gamma_+^n} &= \mel{T\gamma_+^n}{\tau_3}{T\gamma_-^n} \notag \\
    &=-\mel{\gamma_-^n}{\tau_3}{\gamma_+^n} = 0 .
\end{align}
Using the matrix elements found above, we will now write Eq.~\eqref{eq:densitymod} at energies below the superconducting gap, in terms of the surface states. We write the surface Hamiltonians at each TRIM point $\mathbf{k}^\parallel_n$ in terms of $\delta \mathbf{k}^\parallel = \mathbf{k}^\parallel-\mathbf{k}^\parallel_n$ as
\begin{equation}
    H_{n,\mathrm{eff}}(\delta\mathbf{k}^\parallel) = \sum_{ab}\ket{\gamma_a^n}[\vec d_n(\delta\mathbf{k}^{\parallel}) \cdot \vec \Gamma]_{a b}\bra{\gamma_{b}^n} ,\
\end{equation}
where $\vec \Gamma$ is the vector of Pauli matrices in the $\ket{\gamma^n_\pm}$ basis and $\vec d_n(\delta\mathbf{k}^\parallel)$ parametrizes the surface Hamiltonian.
The $\Gamma_i$ basis corresponds to a rotated version of the $\eta_i$ basis used in Ref.~\onlinecite{Christiansen}. The chiral symmetry means that the $z$-component of the $\vec d$-vector vanishes. We write the Green's function in the vicinity of $\k_n^\parallel$ in terms of the positive energy eigenstate of $H_{n,\mathrm{eff}}$ as 
\begin{align}\label{eq:gftrim}
    G_n(\delta\mathbf{k}^\parallel,\omega) 
    &= \frac{1}{2}\sum_{ab}\ket{\gamma_a^n}\frac{[\Gamma_0 +\hat d(\delta \k^\parallel)\cdot \vec \Gamma]_{ab}}{\omega + i\eta -E_{\delta\k^\parallel}}\bra{\gamma_b^n},\
\end{align}
where $\hat d = \vec d/\vert{\vec d}\vert$ and we consider $\omega>0$ so we can neglect the term corresponding to the negative energy eigenstate. Since $\sigma_x$ and $\sigma_z$ anti-commute with $\mathcal U(\mathcal C)$, we can choose the phases of $\ket{\gamma_a^n}$ such that $\mel{\gamma_a^n}{\sigma_{x/z}}{\gamma_b^n}=\zeta_{x/y}[\Gamma_{x/y}]_{ab}$ where $\zeta_{x/y}=\pm 1$. From this, we see that 
\begin{align}
    -\mathrm{Im}\Tr{\sigma_{x}G_n(\delta\mathbf{k}^\parallel,\omega)}= \frac{\eta \zeta_x \hat d_x}{\eta^2 + (\omega-E_{\delta \k^\parallel})^2}, \notag \\
    -\mathrm{Im}\Tr{\sigma_{z}G_n(\delta\mathbf{k}^\parallel,\omega)}= \frac{\eta \zeta_y \hat d_y}{\eta^2 + (\omega-E_{\delta \k^\parallel})^2},
\end{align}
so the $\vec d$-vector can be read off directly from the Green's function. Additionally, $\sigma_y$ commutes with $\mathcal U(\mathcal C)$ and is traceless so $\mel{\gamma_a^n}{\sigma_{y}}{\gamma_b^n}=\zeta_z[\Gamma_{z}]_{ab}$ and $-\mathrm{Im}\Tr{\sigma_yG_n}=0$. This does not break the spin symmetry since $\sigma_i$ do not correspond to physical observables.
\begin{figure}[tb]
    \centering
    \includegraphics[width=\linewidth]{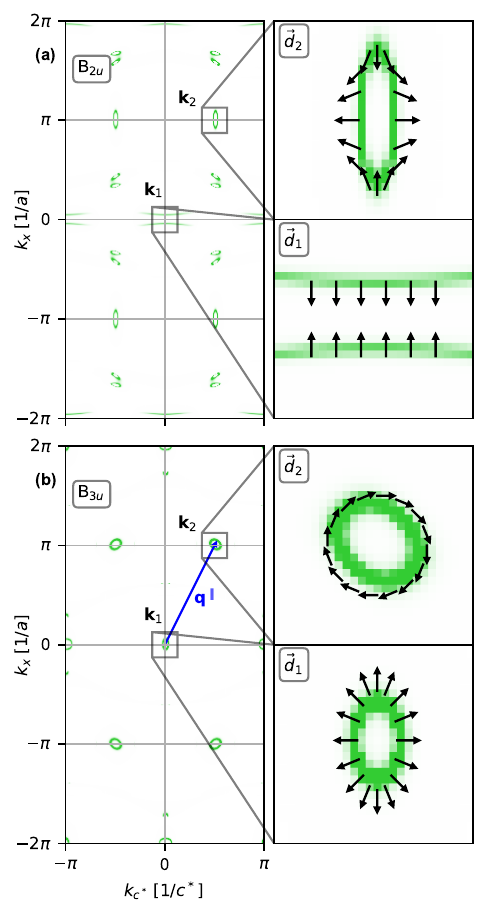}
    \caption{The $\vec d$-vector of the Dirac-cones at $\mathbf{k}_1^\parallel=(0,0)$ and $\mathbf{k}_2^\parallel=(\pi/2,\pi)$ in the $B_{\rm 2u}$ and $B_{\rm 3u}$ phases.}
    \label{fig:surfacedvector}
\end{figure}
\begin{figure}
    \centering
    \includegraphics[width=\linewidth]{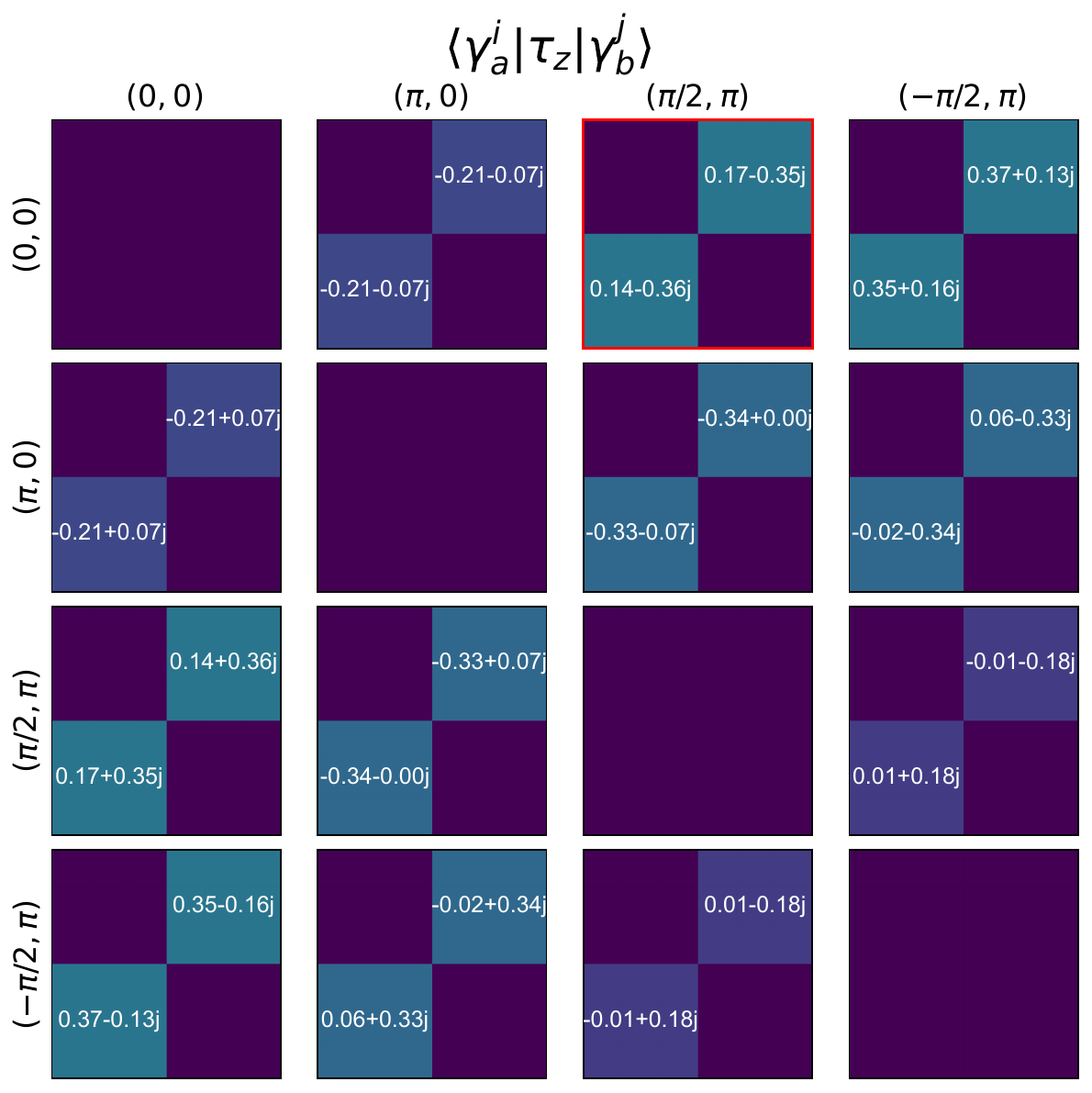}
    \caption{Matrix elements $\mel{\gamma^n_a}{\tau_3}{\gamma_b^m}$, between the basis vectors at TRIM points $\k^\parallel_n$ and $\k^\parallel_m$, from which the overlap amplitude $A$ and phase shift $\eta$ can be read off. }
    \label{fig:tz_mels}
\end{figure}

We now consider the TRIM points where the dispersion is approximately circular, so we can parametrize the $\vec d$-vector as $\vec d_n(\delta\mathbf{k}^\parallel) = E_{\delta\mathbf{k}^\parallel}(\cos \varphi_n,\sin \varphi_n,0)$, where $E_{\delta\mathbf{k}^\parallel} = \abs{\vec{d}(\delta\mathbf{k}^\parallel)}=\nu_n \abs{\delta\mathbf{k}^\parallel}$ and $\nu_n$ is the Fermi velocity of the Dirac cone at $\mathbf{k}_n^\parallel$. Writing $\delta\mathbf{k}^\parallel =r(\cos\theta,\sin\theta)$, the phase $\varphi_n=\varphi_n(\theta)$ depends only on the angle $\theta$. The eigenstates with positive energy are then given by
\begin{equation}\label{eq:gammap}
    \ket{\gamma^n(\varphi_n)} =  \frac{1}{\sqrt{2}}\left(\mathrm{e}^{-i\varphi_n}\ket{\gamma^n_+}+\ket{\gamma_-^n} \right).
\end{equation}
At energies lower than the superconducting gap, $\omega < \Delta_0$, we calculate the density modulations Eq.~\eqref{eq:densitymod} using our expansion as
\begin{figure}
    \centering
    \includegraphics[width=\linewidth]{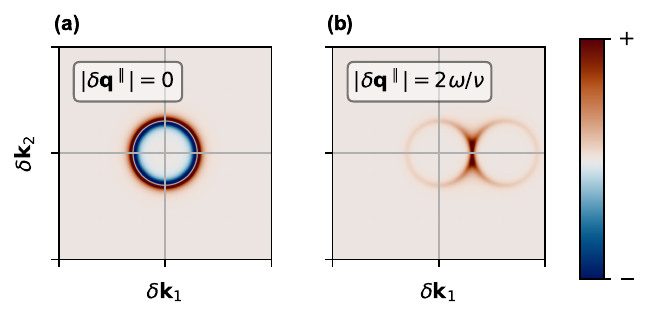}
    \caption{$\delta\rho^{nm}_0$ for $\vert\delta\q^\parallel\vert=0$ and $\vert\delta\q^\parallel\vert=2\omega/\nu$, when $\nu_n=\nu_m$.}
    \label{fig:delta_rho_0}
\end{figure}
\begin{align}\label{eq:densitymod0}
    &\delta \rho_\beta(\mathbf{q}^\parallel,\omega) = -\frac{V_0}{2\pi i}\sum_{\mathbf{k}^\parallel}\bigl(  \Tr{\tau_e G(\mathbf{k}^\parallel+\mathbf{q}^\parallel,\omega)\tau_3 P_\beta G(\mathbf{k}^\parallel,\omega)} \notag \\
    & -\Tr{\tau_e G^\dag(\mathbf{k}^\parallel+\mathbf{q}^\parallel,\omega)\tau_3 P_\beta G^\dag(\mathbf{k}^\parallel,\omega)}\bigr) .
\end{align}
This will be dominated by scattering of surface states between different TRIM points, i.e. at $\mathbf{q}^\parallel\approx \mathbf{k}_n^\parallel-\mathbf{k}_m^\parallel$ for $n\neq m$, explaining the vanishing signal at $\mathbf{q}^\parallel\approx (0,0)$. Expanding $\mathbf{q}^\parallel = \delta\mathbf{q}^\parallel + \mathbf{k}_n^\parallel-\mathbf{k}_m^\parallel$, and using Eq.~\eqref{eq:gftrim}, the contribution to $\delta\rho(\mathbf{q}^\parallel,\omega)$ coming from scattering between $\mathbf{k}^\parallel_n$ and $\mathbf{k}^\parallel_m$ can be written, in terms of $\delta\mathbf{k}^\parallel = r(\cos\theta,\sin\theta)$ and $\delta\mathbf{q}^\parallel+\delta\mathbf{k}^\parallel=r'(\cos\theta',\sin\theta')$, as
\begin{align}
    \delta\rho^{nm}=\sum_{\delta \mathbf{k}^\parallel}\delta \rho_0^{nm}(r,r')
    \Tr{\tau_e P^n(\theta')\tau_3 P_\beta P^m(\theta)}, \notag \\
\end{align}
where $P^n(\theta')=\ket{\gamma^n(\varphi_n)}\bra{\gamma^n(\varphi_n)}$ is the projector onto the surface state $\ket{\gamma^n(\varphi_n)}$ at the TRIM point $\mathbf{k}^\parallel_n$, and $\delta\rho_0^{nm}(r,r')$ is given by
\begin{align}
    \delta \rho_0^{nm}(r,r') &=-\frac{V_0}{\pi }{\rm Im}\qty{\frac{1}{\omega + i\eta -\nu_n r'}\frac{1}{\omega + i\eta -\nu_m r}} \notag \\
    &= -\frac{V_0}{\pi }\frac{\eta\left(2\omega - \nu_nr'-\nu_m r\right)}{\left(\eta^2+(\omega-\nu_n r')^2 \right)\left(\eta^2+(\omega-\nu_m r)^2 \right)} .
\end{align}
When $\delta\mathbf{q}^\parallel=(0,0)$, so $r=r'$, $\delta\rho_0^{nm}(r,r)$ is odd in $r$ leading to a vanishing contribution at all angles when summing over $r$, which corresponds to integrating radially over the integrand shown in Fig. \ref{fig:delta_rho_0}(a). At finite $\delta\mathbf{q}^\parallel$ and assuming that the slope of the Dirac cones is equal, i.e. $\nu_n=\nu$, the main contribution comes from $\abs{\delta\mathbf{q}^\parallel}\approx 2\omega/\nu$, corresponding to scattering between states at opposite sides of the Dirac cone where the contribution is maximal and the integrand is only positive, Fig. \ref{fig:delta_rho_0}(b). At these points, the density modulation can be written as
\begin{align}
    \delta\rho^{nm} \propto& \delta\rho^{nm}_0(r,r') \mel{\gamma^m(\varphi_m(\theta+\pi))}{\tau_e}{\gamma^n(\varphi_n(\theta))} \notag \\
    &\times\mel{\gamma^n(\varphi_n(\theta))}{\tau_3 P_\beta}{\gamma^m(\varphi_m(\theta+\pi)}.
\end{align}
Since $\tau_3P_\beta$ anti-commutes with the chiral symmetry, we can write its matrix elements as
\begin{equation}
    \mel{\gamma_a^n(\varphi^n(\theta))}{\tau_3 P_\beta}{\gamma_b^m(\varphi^m(\theta+\pi)} = A\mqty( && \mathrm{e}^{i\eta} \\ \mathrm{e}^{-i\eta} && )_{ab} ,
\end{equation}
where $\abs{A}=\abs{\mel{\gamma^n_+}{\tau_3}{\gamma^m_-}}$ is the effective overlap amplitude. Using Eq.~\eqref{eq:gammap}, we then find 
\begin{align}\label{eq:tau3overlap}
    &\mel{\gamma^n(\varphi^n(\theta))}{\tau_3 P_\beta}{\gamma^m(\varphi^m(\theta+\pi))} \notag \\ 
    &= A\qty[\mathrm{e}^{i\varphi_n(\theta)-i\eta}+\mathrm{e}^{-i\varphi_m(\theta+\pi)+i\eta}] 
    \notag \\
    &=A\qty[\mathrm{e}^{i\varphi_n(\theta)-i\eta}-\mathrm{e}^{-i\varphi_m(\theta)+i\eta}],
\end{align}
where, in the second equality, we used that $\varphi_n(\theta+\pi) = \varphi_n(\theta)+\pi$ due to the odd-parity of the $\vec d$-vector. From Eq.~\eqref{eq:tau3overlap} we see that the QPI signal vanishes when $\varphi_n(\theta)+\varphi_m(\theta)=2\eta$ modulo $2\pi$. 

As an example, we now consider in detail the QPI signal at $\mathbf{q}^\parallel \approx (\pi/2,\pi)$ in the $B_{\rm 3u}$ phase. We read off the matrix elements $\langle {\gamma^i_a}|{\tau_3P_{\rm Te}}|{\gamma^j_b}\rangle$ from Fig.~\ref{fig:tz_mels} (element with red boundary) and find that $\eta\approx0$ for the $(0,0)$ to $(\pi/2,\pi)$ scattering since the matrix element is approximately proportional to $\sigma_x$. From the $\vec d$-vectors plotted in Fig.~\ref{fig:surfacedvector} we see that $\varphi^{(0,0)}(\theta)+\varphi^{(\pi/2,\pi)}(\theta)\approx 0$ at $\theta= \pi/4$. From this it follows that the nodes in the QPI signal should appear at $\theta = \pi/4,5\pi/4$ while the maxima appear at $\theta = 3\pi/4,7\pi/4$, consistent with Fig.~3(f) in the main text. We find that in the $B_{\rm 3u}$ phase, all the Dirac cones have positive winding number, that is, $\varphi_n = \varphi_{0,n}+\theta$, for some constant phase shift $\varphi_{0,n}$. This means that all QPI peaks at the TRIM points should have intensity nodes, consistent with Fig.~\ref{fig:b3usurfaceqpi}. The calculations developed above do also describe the the QPI signal from the Dirac surface states in Fig.~\ref{fig:qpi_alpha05}(f). However, the presence of additional bulk nodes allows additional signal at low energies from the bulk and further complicates the interpretation of the signal. Apart of such complications, the analysis in this section explains the main qualitative features of topological surface state QPI, and point to new signatures that can be searched for experimentally.

\subsection{U and Te Impurities}

\begin{figure*}
    \centering
    \includegraphics[width=\linewidth]{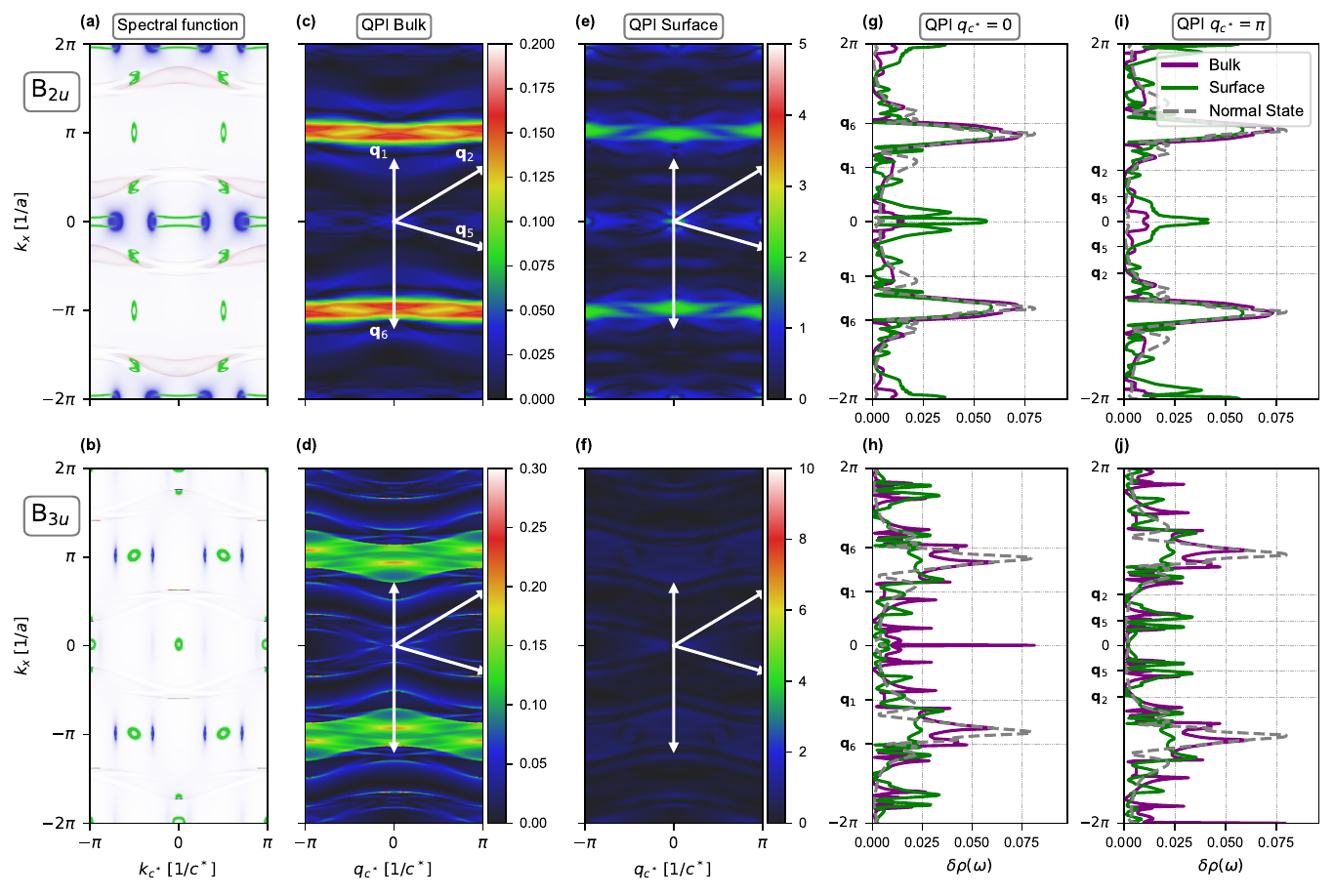}
    \caption{QPI signal and spectral functions on the (0-11) surface of UTe$_2$ in the B$_{2u}$ and B$_{3u}$ phases with $\alpha=0$, and scattering originating only from impurities at the U sites. (a-b) The bulk U (red) and Te (blue) spectral functions along with the surface states (green). (c,d) [e,f] QPI versus surface momentum at $\omega=0.05\Delta_0$ including the surface-projected bulk [surface] states. Characteristic scattering vectors $\mathbf q_i$ are indicated by arrows. (g,h) [(i,j)] Momentum cuts of the respective panels (c,d) [e,f] at $q_{c^*}=0, \pi$. In panels (g-j) we have also indicated the locations of $\mathbf q_i$.}
    \label{fig:qpi_u}
\end{figure*}

\begin{figure*}
    \centering
    \includegraphics[width=\linewidth]{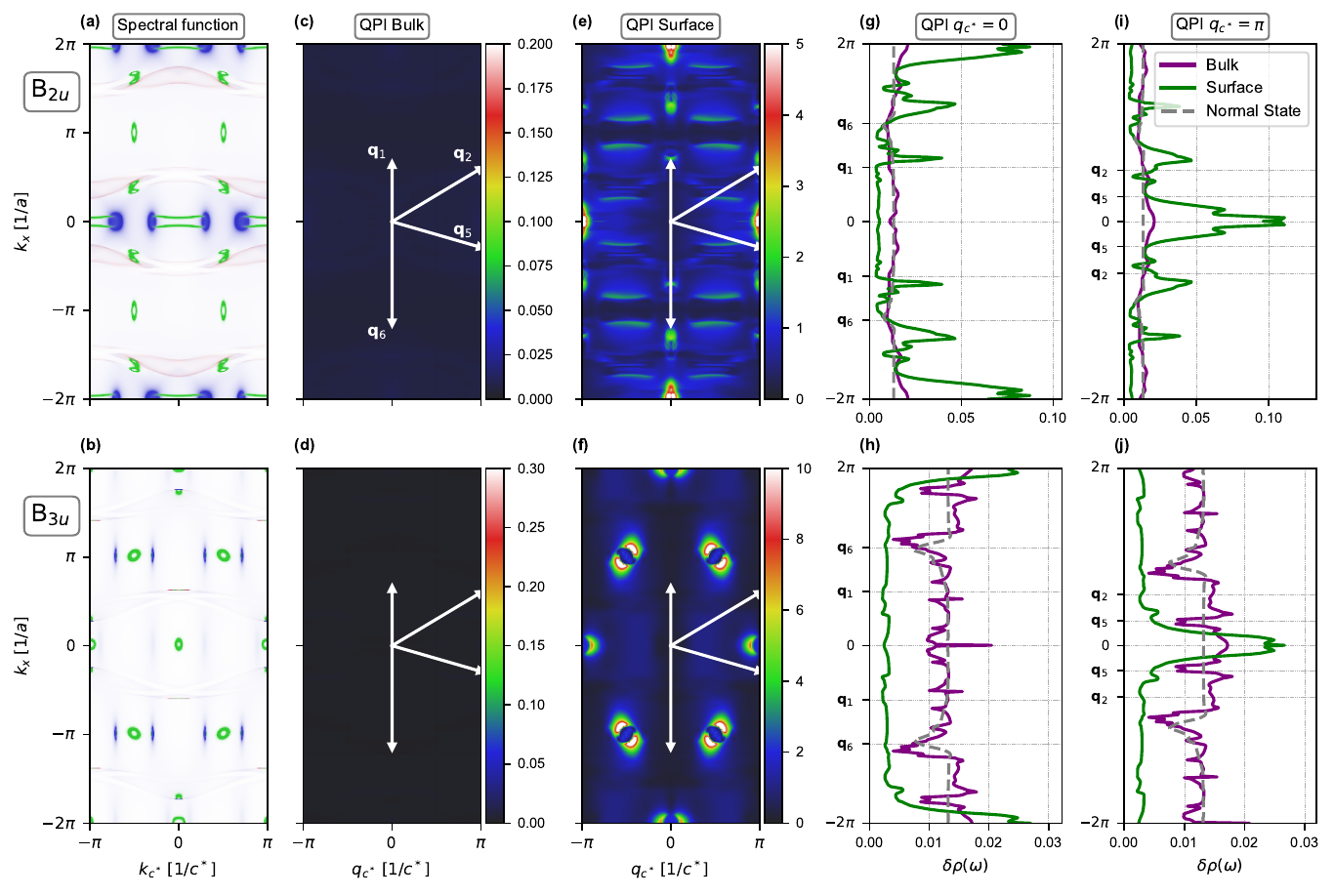}
    \caption{QPI signal and spectral functions on the (0-11) surface of UTe$_2$ in the B$_{2u}$ and B$_{3u}$ phases with $\alpha=0$, and scattering originating only from impurities at the Te sites. (a-b) The bulk U (red) and Te (blue) spectral functions along with the surface states (green). (c,d) [e,f] QPI versus surface momentum at $\omega=0.05\Delta_0$ including the surface-projected bulk [surface] states. Characteristic scattering vectors $\mathbf q_i$ are indicated by arrows. (g,h) [(i,j)] Momentum cuts of the respective panels (c,d) [e,f] at $q_{c^*}=0, \pi$. In panels (g-j) we have also indicated the locations of $\mathbf q_i$.}
    \label{fig:qpi_te}
\end{figure*}

To supplement Fig. 3 in the main text, we show in Figs.~\ref{fig:qpi_u} (\ref{fig:qpi_te}), the QPI signal coming from impurities on the U (Te) sites only. With U impurities, the surface signal (see Fig.~\ref{fig:qpi_u}(e-f)) is highly suppressed since the TSS have predominant weight on the Te sites, while the bulk QPI signal (Fig.~\ref{fig:qpi_u}(c-d)) remains largely unchanged as it mainly arises from the scattering of the dispersing U bands. This should be contrasted to Fig.~\ref{fig:qpi_te} where the bulk signal is almost completely vanished, while the surface signal is almost identical to the Fig. 3(e-f) in the main text. This means that the low energy QPI signal with Te impurities reveal information about the TSS while the U impurities reveal information about the bulk band structure and the quasiparticle nodes of the relevant SC order parameter.

\bibliography{Refs}